%% file: main.tex
\documentclass[compsoc,conference,a4paper,10pt,times,final]{IEEEtran}
\IEEEoverridecommandlockouts

\input{stuff/preambel.tex}
\input{stuff/macros.tex}
\input{stuff/glossaries.tex}

\begin{document}

\title{\Large \bf \glsentrytext{ssomon}: Fully-Automatic Large-Scale Landscape, Security, and Privacy Analyses of \glsentrylong{sso} in the Wild}

\input{stuff/authors_ieee.tex}

\maketitle

\glsresetall
\input{stuff/glsunset.tex}

\input{sections/00_abstract.tex}

\glsresetall

\input{stuff/glsunset.tex}
\input{sections/10_introduction.tex}

\input{sections/20_basics.tex}

\input{sections/25_sso_tools.tex}

\input{sections/30_sso_monitor}
\input{sections/40_ground_truth}
\input{sections/50_landscape}

\input{sections/60_security}

\input{sections/70_privacy}

\input{sections/80_related_work}

\input{sections/90_futurework.tex}

\bibliographystyle{plainnat}
\bibliography{bib/refs,bib/rfc}

\appendix
\input{sections/100_flowchart.tex}
\input{sections/120_attack_list.tex}

\end{document}

%% file: stuff/preambel.tex
\usepackage{xcolor}
\usepackage{graphicx}
\usepackage{amsmath,amssymb,amsfonts,wasysym}
\usepackage{fontawesome5}
\usepackage{hyperref}
\usepackage{url}
\usepackage[numbers]{natbib}
\usepackage{etoolbox}
\usepackage{paralist}
\usepackage[inline]{enumitem}
\usepackage{booktabs}
\usepackage{tabularx}
\usepackage{array}
\usepackage{multirow}
\usepackage{multicol}
\usepackage{subcaption}
\usepackage{comment}
\usepackage[frozencache]{minted}

\definecolor{lightgray}{gray}{0.9}
\definecolor{fsupGreen}{RGB}{40,125,58}

\newcolumntype{Y}{>{\centering\arraybackslash}X}

\setdefaultenum{(1)}{(a)}{(i)}{(A)}

\setminted{
    frame=single, %
    fontsize=\scriptsize,
    linenos=true,
    numbersep=5pt,
    breaklines=true,
    breakautoindent=true,
    breakbefore=.,
    breakafter=-
}
\setmintedinline{
    fontsize=\small,
    escapeinside=~~,
    breakbefore=.,
    breakafter=-:/\,
}
\newmintinline[code]{text}{}
\newmintinline[js]{javascript}{}
\newmintinline[html]{html}{}
\newmintinline[json]{json}{}

\def\Snospace~{\S{}}

\usepackage{titlesec}
\titlespacing*{\paragraph}{\parindent}{0.2ex plus .1ex minus .1ex}{1.5ex} %

\makeatletter
\DeclareRobustCommand{\textsupsub}[2]{{%
  \m@th\ensuremath{%
    ^{\mbox{\fontsize\sf@size\z@#1}}%
    _{\mbox{\fontsize\sf@size\z@#2}}%
  }%
}}
\makeatother

%% file: stuff/macros.tex
\newcommand{\fsup}{\textcolor{fsupGreen}{\faIcon[solid]{check-circle}}} %
\newcommand{\nsup}{\textcolor{red}{\faIcon[regular]{minus-circle}}} %

\newcommand{\num}[1]{\textcolor{black}{#1}} %

\newcommand{\removed}[1]{} %

%% file: stuff/glossaries.tex
\usepackage{glossaries}
\glsdisablehyper
\newcommand{\glssc}[1]{\textsc{\glsentryuseri{#1}}}

\newglossaryentry{ssomon}{name={\textsc{SSO-Monitor}},description={},user1={SSO-Monitor}}
\newglossaryentry{spar}{name={\textsc{SmartParams}},description={},user1={SmartParams},user2={\textsc{SmartParam}}}

\newacronym{oauth}{OAuth}{OAuth Authorization Framework 2.0}
\newacronym{oidc}{OIDC}{OpenID Connect 1.0}

\newglossaryentry{user}{name=user,description=}
\newacronym{ua}{UA}{User Agent}
\newacronym{client}{Client}{Client Application}
\newacronym{idp}{IdP}{Identity Provider}

\newglossaryentry{loginreq}{name=\emph{login request},description=}
\newglossaryentry{loginresp}{name=\emph{login response},description=}
\newglossaryentry{tokenreq}{name=\emph{token request},description=}
\newglossaryentry{tokenresp}{name=\emph{token response},description=}

\newacronym{httpc}{HttpCom}{HTTP Communication}
\newacronym{ibc}{InBC}{In-Browser Communication}
\newacronym{sdk}{SDK}{Software Development Kit}

\newglossaryentry{code}{name=\texttt{code},description=}
\newglossaryentry{at}{name=\texttt{access\_token},description=}
\newglossaryentry{idt}{name=\texttt{id\_token},description=}

\newacronym{dfs}{DFS}{Depth-First Search}
\newacronym{bfs}{BFS}{Breath-First Search}

\newacronym{http}{HTTP}{Hypertext Transfer Protocol}

\newacronym[]{saas}{SaaS}{Software-as-a-Service}
\newacronym[]{iaas}{IaaS}{Infrastructure-as-a-Service}
\newacronym[]{paas}{PaaS}{Platform-as-a-Service}
\newacronym[]{fin}{\emph{FIN}}{Finished message}
\newacronym[]{tls}{TLS}{Transport Layer Security}
\newacronym[]{tlsu}{TLS Unique}{TLS Unique}
\newacronym[]{sessc}{session cookie}{HTTP Session Cookie}
\newacronym[]{hmac}{HMAC}{Keyed-Hash Message Authentication Code}
\newacronym[]{hok}{HoK}{Holder-of-Key}
\newacronym[]{xref}{XRef}{Cross Reference Table}

\newacronym[]{ca}{CA}{Certificate Authority}
\newacronym{fido}{FIDO}{Fast IDentity Online}
\newacronym{poc}{PoC}{Proof-Of-Concept}
\newacronym[]{opiv}{PrOfESSOS}{Practical Offensive Evaluation of Single Sign-On Services}
\newacronym[]{str}{STR}{Security Test Runner}
\newacronym[]{eaas}{EaaS}{Evaluation-as-a-Service}
\newacronym[]{att}{AT}{Attacker Type}
\newacronym[]{trc}{TRC}{Token Recipient Confusion}
\newacronym[]{kc}{KC}{Key Confusion}
\newacronym[]{ids}{IDS}{ID Spoofing}
\newacronym[]{ds}{DS}{Discovery Spoofing}
\newglossaryentry{wr}{name=Wrong Recipient,description=}
\newglossaryentry{sb}{name=Signature Bypass,plural=Signature Bypasses,description=}
\newglossaryentry{r}{name=Replay,description=}
\newglossaryentry{ra}{name=Replay Attack,description=}
\newglossaryentry{cr}{name=Covert Redirect,description=}
\newglossaryentry{ssot}{name=authentication token,description=}
\newglossaryentry{c}{name=Client,description=}
\newglossaryentry{upw}{name={Username/Password}, description={}}
\newglossaryentry{wb}{%
    name={web browser},
    plural={web browsers},
    description={},
}
\newacronym[]{idm}{IdM}{Identity Management}
\newacronym[%
description={\glsentrylong{aaas}; A SOA concept which provides an authentication service as a service},
]
{aaas}{AaaS}{authentication as a service}
\newacronym[%
description={\glsentrylong{ax}; OpenID extension for exchanging identity information between endpoints~\cite{ax}},
]
{ax}{Ax}{Attribute Exchange}
\newacronym[%
description={\glsentrylong{sreg}; OpenID extension for exchanging identity information between endpoints~\cite{sreg}},
]
{sreg}{SReg}{Simple Registration}
\newacronym[%
description={\glsentrylong{cms}; A software which helps to create an manage content collaboratory. CMSs are often web applications},
]
{cms}{CMS}{content-management system}
\newacronym[%
description={\glsentrylong{dos}; The unavailability of a service, which should be available},
]
{dos}{DoS}{Denial-of-Service}
\newacronym[%
description={\glsentrylong{gui}; A software component which allows a human to interact with machine by using symbols},
]
{gui}{GUI}{Graphical User Interface}
\newacronym[%
description={\glsentrylong{jaxb}; An interface for Java allowing to bind XML Data to Java objects. See \url{https://jaxb.java.net/}},
]
{jaxb}{JAXB}{Java Architecture for XML Binding}
\newacronym[%
description={\glsentrylong{mitm}; An attack technique in which the attacker virtually or physically sits between to users, allowed to listen and manipulate their exchanged messages},
]
{mitm}{MitM}{Man-in-the-Middle}
\newacronym[
description={Single Sign-On is a property of access controll, which allows a user to log in once and request access to several systems},
]{sso}{SSO}{Single Sign-On}
\newacronym[%
description={\glsentrylong{soa}; Abstract model of software architecture},
]
{soa}{SOA}{Service Oriented Architecture}
\newacronym[%
description={\glsentrylong{sp}; Also known as \emph{Relying Party}, entity that offers a service which the user wants to use},
]
{sp}{SP}{Service Provider}
\newacronym[%
description={\glsentrylong{w3c}; Organization for standards in the World Wide Web},
]
{w3c}{W3C}{World Wide Web Consortium}
\newacronym[%
description={\glsentrylong{xml}; Textual data format to encode documents, commonly used for message exchange~\cite{xml}},
]
{xml}{XML}{eXtended Markup Language}
\newacronym[%
description={\glsentrylong{xsw}; Technique for attacking signed XML documents~\cite{McIntosh2005}},
]
{xsw}{XSW}{XML Signature Wrapping}
\newacronym[%
description={\glsentrylong{har}; A JSON-based archive format for HTTP traffic},
]
{har}{HAR}{HTTP Archive format}

\newacronym[]{certfaking}{CF}{Certificate Faking}
\newacronym[]{sigexclusion}{$\emptyset$Sig}{Signature Exclusion}
\newacronym[]{certinjection}{CInj}{Certificate Injection}
\newacronym[]{xxeattack}{XXEA}{XML External Entity Attack}
\newacronym[]{xsltattack}{XSLTA}{XSLT Attack}
\newacronym[]{xslt}{XSLT}{Extensible Stylesheet Language Transformation}
\newacronym[]{acsspoofing}{ACS Spoofing}{\emph{AssertionConsumerServiceURL} Spoofing}
\newacronym[]{pop}{PoP}{Proof-of-Possession}

\newacronym{fapi}{FAPI}{Financial-grade API}

\newglossaryentry{acsscanner}{%
    name={ACS-Scanner},
    description={A penetration testing tool used to discover ACS Spoofing vulnerabilities at an IdP.},
}
\newglossaryentry{samlattacker}{%
    name={SAML-Attacker},
    description={A penetration testing tool used to discover SAML based vulnerabilities at an SP.},
}
\newglossaryentry{xss}{%
    name={XSS},
    description={Cross-Site-Scripting is a web application vulnerability which tries to execute an attacker script within the context of the website owner within the user's browser},
}
\newglossaryentry{bf}{%
    name={Brute-Force},
    description={Brute-Force}
}

\newglossaryentry{pbl}{%
    name={password-based login},
    description={password-based login}
}
\newglossaryentry{PBL}{%
    name={Password-Based Login},
    description={password-based login}
}

\newglossaryentry{authreq}{%
    name={\emph{AuthnReq}},
    description={Authentication Request initiating the \gls{sso} authentication scheme. It is usually issued by the \gls{sp} and send to the \gls{idp}},
}
\newglossaryentry{authresp}{%
    name={\emph{AuthnRes}},
    description={Authentication Response containing the information about the authenticated \glsentry{enduser}. It is usually issued by the \gls{idp} and send to the \gls{sp}},
}
\newglossaryentry{clientid}{name={\texttt{client\_id}},description={Unique string identifying the \gls{sp}.}}
\newglossaryentry{clientsecret}{name={\texttt{client\_secret}},description={A shared secret between the IdP and SP used for authentication of the SP.}}
\newglossaryentry{redirecturi}{name={\texttt{redirect\_uri}},description={A callback URI of the \gls{sp} where the authentication token is sent by the IdP.}}

\newglossaryentry{dhke}{%
	name={Diffie-Hellman key exchange},
	description={Specific method for exchanging key material},
}
\newglossaryentry{asso}{%
	name={association},
	description={An association between the SP and the OpenID IdP establishes a shared secret between them~\cite[Section 8]{openid20}},
}
\newacronym{csrf}{CSRF}{Cross-Site-Request-Forgery}
\newglossaryentry{dirver}{%
	name={direct verification},
	description={OpenID signature verfication performed by the OpenID IdP itself, enforced by the SP~\cite[Section 11.4.2]{openid20}},
}
\newglossaryentry{junit}{%
	name={JUnit},
	description={JUnit is framework for testing Java programs},
}
\newglossaryentry{testng}{%
	name={TestNG},
	description={TestNG is framework for testing Java programs},
}
\newglossaryentry{mysql}{%
	name={MySQL},
	description={MySQL is a widely used \glsentrytext{opensource} relational database management system},
}
\newglossaryentry{oid}{%
    name={OpenID},
    description={OpenID is decentralized authentication system for web-based SPs~\cite{openid20}},
}
\newglossaryentry{oid4j}{%
	name={OpenID4Java},
	description={OpenID4Java is library which easily allows to implement an OpenID consumber as well as an OpenID IdP ~\cite{openid4java}},
}
\newglossaryentry{oida}{%
	name={OpenID Attacker},
	description={OpenID Attacker is the tool developed in this thesis for attacking the OpenID specification},
}
\newglossaryentry{opensource}{%
	name={open source},
	description={Work which are enforced by license to have source available for the public},
}
\newglossaryentry{oc}{
	name={ownCloud},
	description={OwnCloud is a free and \glsentrytext{opensource}, \glsentrytext{php} and \glsentrytext{mysql} based web application which allows data synchronization and cloud storage~\cite{owncloud}.}
}
\newglossaryentry{pentest}{%
	name={penetration test},
	user1={Penetration Test},
	user2={Penetration Tests},
	user3={penetration testing},
	user4={Penetration Testing},
	description={Method for evaluating security on computer systems},
}
\newglossaryentry{pentester}{%
	name={penetration tester},
	description={A person evaluating security on computer systems},
}
\newglossaryentry{php}{%
	name={PHP},
	description={PHP is a popular server-side scripting language and widespread in the area of web development},
}
\newglossaryentry{saml}{%
	name={SAML},
	description={Security Assertion Markup Language~\cite{saml}},
}
\newglossaryentry{browserid}{%
	name={BrowserId},
	description={},
}
\newglossaryentry{sqlinjection}{
	name={SQL-Injection},
	description={An attack technique which tries to inject and execute SQL statements reasoned in inadequate user input validation~\cite{sqlinjection}},
}
\newglossaryentry{wsattacker}{%
	name={WS-Attacker},
	description={Automatic \glsentrytext{pentest} framework~\cite{wsattacker}},
}
\newglossaryentry{doc}{%
	name={XML document},
	description={General term for a document which uses \glsentrytext{xml} as description language, e.g.\ a \glsentrytext{soap} message},
	parent=xml,
	user1={document},
	user2={documents},
}
\newglossaryentry{wa}{%
	name={web application},
	user1={Web Application},
	user2={Web Applications},
	description={A web application is an application that uses a web browser as a client.}
}
\newglossaryentry{wp}{%
	name={WordPress},
	description={WordPress is a free and \glsentrytext{opensource}, \glsentrytext{php} and \glsentrytext{mysql} based \glsentrytext{cms}~\cite{wordpress}.}
}
\newglossaryentry{xmlsig}{%
	name={XML Signature},
	user1={XML Digital Signature},
	description={also \glsentryuseri{xmlsig}; Standard for creating signatures in \glsentrytext{xml} documents~\cite{xmlsig1, xmlsig2}},
}
\newglossaryentry{xrds}{%
	name={XRDS},
	description={eXtensible Resource Descriptor Sequence is an XML format for describing metadata as a web resource},
}

\newglossaryentry{enduser}{%
	name={user},
	description={A user surfing in the web via his User-Agent},
}
\newglossaryentry{idpc}{%
	name={IdP Confusion},
	description={},
}

\newacronym[longplural=Relying Parties]{rp}{RP}{Relying Party}
\newacronym[]{kdc}{KDC}{Key-Distribution Center}
\newacronym[]{https}{HTTPS}{Hypertext Transfer Protocol Secure / HTTP Over TLS}
\newacronym[]{url}{URL}{Uniform Resource Locator}
\newacronym[]{uri}{URI}{Uniform Resource Identifier}
\newacronym[]{api}{API}{Application Programming Interface}
\newacronym[]{rest}{REST}{Representational State Transfer}
\newacronym[]{json}{JSON}{JavaScript Object Notation}
\newacronym[]{jwt}{JWT}{JSON Web Token}
\newacronym[]{jwks}{JWKS}{JSON Web Key Set}
\newacronym[]{jws}{JWS}{JSON Web Signature}
\newacronym[]{jwe}{JWE}{JSON Web Encryption}
\newacronym[]{ic}{IC}{Issuer Confusion}
\newacronym[]{sm}{SM}{Signature Manipulation}
\newacronym[]{sessm}{SM}{Session Management}
\newacronym[]{aam}{AAM}{Authorization \& Access Management}
\newacronym[]{scs}{SCS}{Sub Claim Spoofing}
\newacronym[]{rum}{RUM}{Redirect URI Manipulation}
\newacronym[]{ietf}{IETF}{Internet Engineering Task Force}
\newacronym[]{arm}{aRM}{Attacker Related Message}
\newacronym[]{brm}{BRM}{Browser Relayed Message}
\newacronym[]{ttp}{TTP}{Trusted Third Party}
\newacronym[]{ta}{TA}{tainted address}
\newacronym[]{ssrf}{SSRF}{Server Side Request Forgery}
\newacronym[]{ac}{\texttt{code}}{Authorization Code}
\newacronym[]{dtd}{DTD}{Document Type Definition}
\newglossaryentry{entity}{%
    name={XML Entity},
    plural={Entities},
    description={Entities are used in DTDs.},
}
\newglossaryentry{exentity}{%
    name={External Entity},
    plural={External Entities},
    description={External Entities are used in DTDs and allow the access of system files or other resources.},
}

\newglossaryentry{oidcp}{%
	name={OpenID Connect protocol},
	description={},
}

\newglossaryentry{dynreg}{%
	name={Dynamic Registration},
	description={A phase in OpenID Connect specifying how an SP registers dynamically on an IdP without any user interaction and configuration.},
}

\newacronym[]{regEndp}{\texttt{regEndp}}{\glsentrytext{dynreg} Endpoint}
\newacronym[]{authEndp}{\texttt{authEndp}}{Authorization Endpoint}
\newacronym[]{tokenEndp}{\texttt{tokenEndp}}{Token Endpoint}
\newacronym[]{userInfoEndp}{\texttt{userInfoEndp}}{Userinfo Endpoint}
\newacronym[]{jwksEndp}{\texttt{jwksEndp}}{JWKS Endpoint}

\newglossaryentry{mep}{name=Malicious Endpoints,description=}
\newglossaryentry{mepa}{name=Malicious Endpoints Attack,description=}
\newglossaryentry{disc}{name=Discovery,description=}
\newglossaryentry{midp}{name=malicious \glsentrytext{idp},description=}
\newglossaryentry{hidp}{name=honest \glsentrytext{idp},description=}
\newglossaryentry{msp}{name=attacker \glsentrytext{sp},description=}
\newglossaryentry{mds}{name=malicious Discovery service,description=}
\newglossaryentry{redirect_uri}{name=\texttt{redirect\_uri},description=}
\newglossaryentry{client_id}{name=\texttt{client\_id},description=}
\newglossaryentry{scope}{name=\texttt{scope},description=}
\newglossaryentry{response_type}{name=\texttt{response\_type},description=}
\newglossaryentry{state}{name=\texttt{state},description=}
\newglossaryentry{grant_type}{name=\texttt{grant\_type},description=}
\newglossaryentry{authorization_code}{name=\texttt{authorization\_code},description=}
\newglossaryentry{client_secret}{name=\texttt{client\_\-secret},description=}
\newglossaryentry{token_type}{name=\texttt{token\_type},description=}
\newglossaryentry{expires_in}{name=\texttt{expires\_in},description=}
\newglossaryentry{refresh_token}{name=\texttt{refresh\_token},description=}
\newglossaryentry{iss}{name=\texttt{issuer},description=}
\newglossaryentry{sub}{name=\texttt{subject},description=}
\newglossaryentry{aud}{name=\texttt{audience},description=}
\newglossaryentry{iat}{name=\texttt{timestamp},description=}
\newglossaryentry{exp}{name=\texttt{expired},description=}
\newglossaryentry{nonce}{name=\texttt{nonce},description=}
\newglossaryentry{id_token}{name=\texttt{id\_token},description=}
\newglossaryentry{access_token}{name=\texttt{access\_token},description=}
\newglossaryentry{token}{name=\texttt{token},description=}
\newglossaryentry{resource}{name=\texttt{resource},description=}
\newglossaryentry{rel}{name=\texttt{rel},description=}
\newglossaryentry{href}{name=\texttt{href},description=}
\newglossaryentry{issuer}{name=\texttt{issuer},description=}
\newglossaryentry{authorization_endpoint}{name=\texttt{authorization\_endpoint},description=}
\newglossaryentry{token_endpoint}{name=\texttt{token\_endpoint},description=}
\newglossaryentry{jwks_uri}{name=\texttt{jwks\_uri},description=}
\newglossaryentry{response_types_supported}{name=\texttt{response\_types\_supported},description=}
\newglossaryentry{subject_types_supported}{name=\texttt{subject\_types\_supported},description=}
\newglossaryentry{id_token_signing_alg_values_supported}{name=\texttt{id\_token\_\-signing\_alg\_values\_supported},description=}
\newglossaryentry{application_type}{name=\texttt{application\_type},description=}
\newglossaryentry{redirect_uris}{name=\texttt{redirect\_uris},description=}
\newglossaryentry{azp}{name=\texttt{azp},description=}
\newglossaryentry{claims}{name=\texttt{claims},description=}
\newglossaryentry{request}{name=\texttt{request},description=}
\newglossaryentry{request_uri}{name=\texttt{request\_uri},description=}
\newglossaryentry{auth_time}{name=\texttt{auth\_time},description=}
\newglossaryentry{acr}{name=\texttt{acr},description=}
\newglossaryentry{client_secret_basic}{name=\texttt{client\_secret\_basic},description=}
\newglossaryentry{client_secret_post}{name=\texttt{client\_secret\_post},description=}
\newglossaryentry{client_secret_jwt}{name=\texttt{client\_secret\_jwt},description=}
\newglossaryentry{private_key_jwt}{name=\texttt{private\_key\_jwt},description=}
\newglossaryentry{claims_parameter_supported}{name=\texttt{claims\_parameter\_supported},description=}
\newglossaryentry{request_parameter_supported}{name=\texttt{request\_parameter\_supported},description=}
\newglossaryentry{at_hash}{name=\texttt{at\_hash},description=}
\newglossaryentry{c_hash}{name=\texttt{c\_hash},description=}

\newglossaryentry{spa}{name=Single-Phase Attack,user2=Single-Phase,description=}
\newglossaryentry{cpa}{name=Cross-Phase Attack,user1=Cross-Phase,description=}

\glsunset{xss}
\newglossaryentry{consentpage}{name={Consent-Page},description={}}
\newacronym{adfs}{ADFS}{Active Directory Federation Service}
\newglossaryentry{fbc}{name={Facebook Connect},description={Monolithic \glsentrytext{sso} protocol created by Facebook},}
\newglossaryentry{msa}{name={Microsoft Account},description={Monolithic \glsentrytext{sso} protocol created by Microsoft.},}
\newacronym[]{gdpr}{GDPR}{General Data Protection Regulation}
\newglossaryentry{testablishment}{name={trust establishment},description={Phase 1 of an \glsentrytext{sso} protocol between \glsentrytext{sp} and \glsentrytext{idp}}}
\newglossaryentry{tgeneration}{name={token generation},description={Phase 2 of an \glsentrytext{sso} protocol between \glsentrytext{enduser} and \glsentrytext{idp}}}
\newglossaryentry{tredemption}{name={token Redemption},description={Phase 3 of an \glsentrytext{sso} protocol between \glsentrytext{sp} and \glsentrytext{idp}}}
\newglossaryentry{lreq}{name={login request},description={\Glsentrytext{sso} message initiated by the \glsentrytext{enduser} in order to start the \glsentrytext{sso} protocol}}
\newglossaryentry{lres}{name={login response},description={}}
\newglossaryentry{treq}{name={token request},description={\Glsentrytext{sso} message initiated by the \glsentrytext{sp} and sent to the \glsentrytext{idp} requesting the token}}
\newglossaryentry{tres}{name={token response},description={\Glsentrytext{sso} message initiated by the \glsentrytext{ido} and sent to the \glsentrytext{idp} containing the token}}
\newglossaryentry{ired}{name={internal HTTP redirect},description={}}

\newglossaryentry{fc}{name={front-channel},description={}}
\newglossaryentry{bc}{name={back-channel},description={}}

\newacronym[]{pkce}{PKCE}{Proof Key for Code Exchange}

\newacronym[]{dom}{DOM}{Document Object Model}

\newacronym{coiu}{COIU}{Cross-Origin Identifier Unlinkability}
\newacronym{fpi}{FPI}{First-Party Isolation}
\newacronym{itp}{ITP}{Intelligent Tracking Prevention}
\newacronym{etp}{ETP}{Enhanced Tracking Prevention}

\newacronym{dpop}{DPoP}{Demonstrating Proof-of-Possession}

\newacronym{svg}{SVG}{Scalable Vector Graphics}

\newacronym{tp}{TP}{True Positive}
\newacronym{tn}{TN}{True Negative}
\newacronym{fp}{FP}{False Positive}
\newacronym{fn}{FN}{False Negative}

\newacronym{owasp}{OWASP}{Open Web Application Security Project}

\newglossaryentry{js}{name={JavaScript},description={}}

\newacronym{mtls}{mTLS}{mutal TLS}
\newglossaryentry{pm}{name={\emph{postMessage}},description={}}
\newglossaryentry{cm}{name={\emph{Channel Messaging}},description={}}

\newglossaryentry{2fa}{name={2FA},description={}}
\newglossaryentry{captcha}{name={CAPTCHA},description={}}
\newglossaryentry{sms}{name={SMS},description={}}

\newacronym{html}{HTML}{Hypertext Markup Language}

%% file: stuff/authors_ieee.tex
\newcommand{\rub}{\textit{Ruhr University Bochum}}
\newcommand{\hshb}{\textit{Heilbronn University of Applied Sciences}}

\author{
    \IEEEauthorblockN{
        Maximilian Westers$^1$,
        Tobias Wich$^2$, 
        Louis Jannett$^2$,
        Vladislav Mladenov$^2$,
        Christian Mainka$^2$,
        Andreas Mayer$^1$
    }
    \IEEEauthorblockA{
        $^1$ \hshb, \\ \texttt{\{maximilian.westers,andreas.mayer\}@hs-heilbronn.de}
    }
    \IEEEauthorblockA{
        $^2$ \rub, \\
        \texttt{\{tobias.wich,louis.jannett,vladislav.mladenov,christian.mainka\}@rub.de}
    }
}

%% file: stuff/glsunset.tex
\glsunset{url}
\glsunset{svg}
\glsunset{ietf}
\glsunset{w3c}
\glsunset{http}
\glsunset{csrf}
\glsunset{owasp}
\glsunset{json}
\glsunset{api}
\glsunset{html}

%% file: sections/00_abstract.tex
\begin{abstract}

\gls{sso} shifts the crucial authentication process on a website to %
to the underlying \gls{sso} protocols and their correct implementation.
To strengthen \gls{sso} security, organizations, such as \gls{ietf} and \gls{w3c}, maintain advisories to address known threats. %
One could assume that these security best practices are widely deployed on websites.
We show that this assumption is a fallacy.  

We present \gls{ssomon}, an open-source fully-automatic large-scale \gls{sso} landscape, security, and privacy analysis tool.
In contrast to all previous work, \gls{ssomon} uses a highly extensible, fully automated workflow with novel visual-based \gls{sso} detection techniques, enhanced security and privacy analyses, and continuously updated monitoring results.
It receives a list of domains as input to discover the login pages, recognize the supported \glspl{idp}, and execute the \gls{sso}.
It further reveals the current security level of \gls{sso} in the wild compared to the security best practices on paper.  

With \gls{ssomon}, we automatically identified \num{1,632} websites with \num{3,020} Apple, Facebook, or Google logins within the Tranco \num{10k}.
Our continuous monitoring also revealed how quickly these numbers change over time.
\gls{ssomon} can automatically login to each \gls{sso} website. %
It records the logins by tracing \gls{http} and in-browser communication to detect widespread security and privacy issues automatically.
We introduce a novel deep-level inspection of \gls{http} parameters that we call \gls{spar}.
Using \gls{spar} for security analyses, we uncovered \gls{url} parameters in \num{5} \gls{client} secret leakages and \num{337} cases with weak \gls{csrf} protection.
We additionally identified \num{447} cases with no \gls{csrf} protection, \num{342} insecure \gls{sso} flows and \num{9} cases with nested URL parameters, leading to an open redirect in one case.
On top, \gls{ssomon} reveals privacy leakages that deanonymize users and allow user tracking without user awareness in \num{200} cases.

\end{abstract}

%% file: sections/10_introduction.tex
\section{Introduction}
\label{sec:intro}

\begin{figure}[t]
    \centering
    \includegraphics[width=\linewidth]{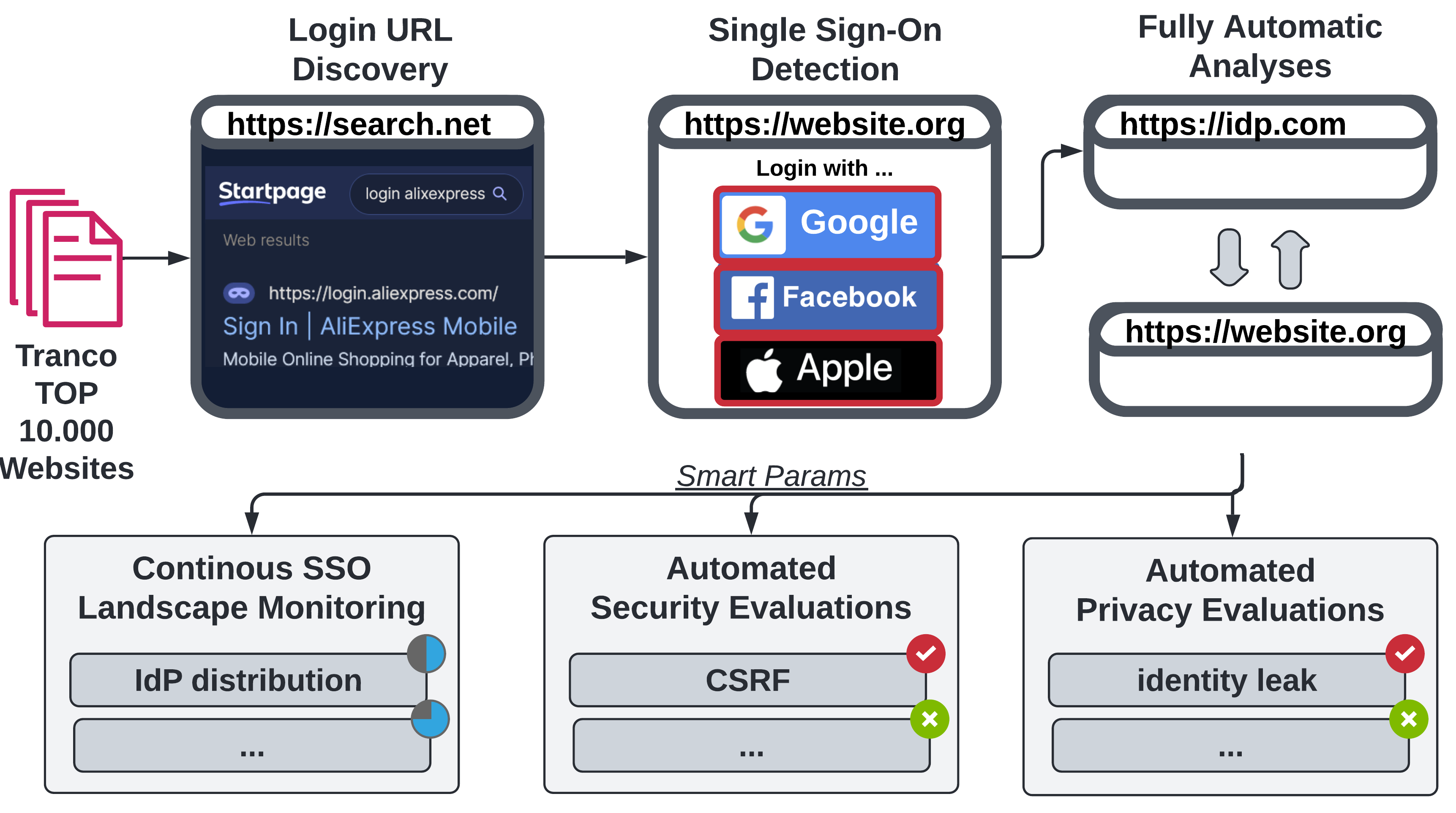}
    \caption{
    \gls{ssomon}'s Fully-Automatic Workflow.
    It starts with the input of a Tranco list and employs search engines to discover a website's login page.
    Then, it uses both a visual-based and pattern-based approach to detect the \glsentryshort{sso} login buttons. %
    It executes \glsentryshort{sso} and conducts security and privacy analyses on the login traces.
    \gls{ssomon} outputs landscape, security, and privacy results.
    }
    \label{fig:intro}
\end{figure}

\Gls{sso} allows websites to quickly register and login users to their accounts by using popular \glspl{idp} like Apple, Facebook, and Google.
The user authentication is provided by implementing two de-facto standards for \gls{sso}: \gls{oauth} and \gls{oidc}.
Both protocols provide a flexible and user-transparent way to share resources, such as profile information, between the website, which acts as a \gls{client}, and an \gls{idp}.
For users, this offers the opportunity to handle only one central account at the \gls{idp}, but still be able to use multiple \glspl{client}.
Authentication flaws are among the \gls{owasp} Top Ten~\cite{owasptop10} vulnerabilities and, as such, of prime importance.
Previously, passwords played a major role in authentication, but since they are known to be problematic~\cite{passwords_herley,passwords_jakobsson,passwords_veras}, \gls{sso} seems to be the promising solution.
However, \gls{sso} is evolving quickly.
For example, among the four proposed OAuth protocol variants (grant types) from 2012~\cite{oauth}, only one (the code grant) is still considered secure if combined with various extensions~\cite{I-D.ietf-oauth-security-topics}.
Developers can hardly follow and implement the recommendations.
To address these issues, researchers investigated \gls{sso} repetitively.

\paragraph{Prior Work}
The majority of related work
~\cite{Wang2012, authscan, blackboxMPWA_NDSS16, yang2016model, sok_oidc, yang2018, rahat2019, li2019oauthguard, wei2021, rahat2022, jannett2022}
implemented \gls{sso} security tools that require manual detection and execution of \gls{sso} logins.
Due to the lack of a fully automated evaluation, researchers often limit the evaluation to a small subset of the most frequently used websites, such as Alexa or Tranco \num{1k} and one \gls{idp}.
This subset makes it hard to estimate how frequently common vulnerabilities appear and the implementation of security best practices are adopted on the Internet.
Other studies automated the \gls{sso} detection and execution but
did not investigate \gls{sso} security and privacy~\cite{authscope, jonker2018shepherd, drakonakis2020cookie}
or only parts of the login flow~\cite{morkonda2021}.
Still, some researchers
~\cite{ssoscan, Shernan2015, ghasemisharifSingleSignOffWhere2018, ghasemisharif2022}
performed large-scale \gls{sso} security and privacy analyses.
However, many results are not reproducible since the used tools are not publicly available.

\paragraph{Challenges in Monitoring SSO.}
Evaluating the security of large-scale \gls{sso}-deployments automatically is a challenging task.
First, it is hard to quickly and reliably determine which websites support \gls{sso} when only its domain is known.
In contrast to previous work, we implemented a novel visual-based \gls{sso} detection for increased accuracy.
Second, large-scale security and privacy evaluations are restricted to passive traffic analyses due to ethical considerations.
To gain more information and thus conduct more comprehensive evaluations, we are the first to introduce a new deep parameter inspection technique (\gls{spar}).
This approach allows for uncovering low-entropy security parameters (CSRF protection), nested parameters inside \gls{http} traffic (secret leakages), and URL parameters (open redirect), which prior work missed.

Our work answers the following three research questions.

\paragraph{RQ1: How can we continuously monitor the \gls{sso} landscape at scale?}
We noticed that the \gls{sso} support on websites varies over time so that prior surveys cannot represent the current \gls{sso} landscape.
Websites are frequently adding new \glspl{idp} and removing others.
The wide deployment of Apple's \gls{idp} proves that introducing new \glspl{idp} can entirely revamp the \gls{sso} landscape in only three years.
Therefore, we see the demand for an automated approach as mandatory before any empirical evaluations of real-world \gls{sso} implementations can be conducted.
Although automatically monitoring the \gls{sso} landscape might seem to be an engineering problem, numerous novel challenges and in-depth research must be conducted.
We designed a modular architecture to solve these challenges:
\begin{inparaenum}
    \item We establish a methodology to automatically find the login page for a given domain with search engines.
    \item We detect the \glspl{idp} that a website supports by recognizing their logos and searching for patterns and keywords.
    \item We automatically execute \gls{sso} logins, including interactions with \glspl{idp} such as consenting.
\end{inparaenum}
For proofing the correctness of our approach, we conducted a \emph{manual} ground truth analysis on the Tranco \num{1k} and compared the results with our automated approach.
We then extended the automated analyses to the Tranco \num{10k} and provide novel insights on the \gls{sso} landscape.
We identified \num{3,020} \gls{sso} logins in the Tranco \num{10k} and provide details on their supported \glspl{idp} and protocol details (see \autoref{sec:landscape}).

\paragraph{RQ2: How secure are current \gls{sso} logins?}
As a result of several discovered vulnerabilities in the recent years, \gls{ietf} created a constantly updated draft addressing all known security issues~\cite{I-D.ietf-oauth-security-topics, RFC6819}.
The \gls{ietf} also created multiple additional documents, such as \glsentryshort{jwt} best practices~\cite{RFC8725}, \glsentryshort{pkce}~\cite{RFC7636}, and \glsentryshort{mtls}~\cite{RFC8705}, to strengthen \gls{sso}.
The question arises if all these security considerations and improvements are implemented to protect the users relying on \gls{sso}.
We systematize the current state of the applied security mechanisms on the Internet. In \num{342} cases \gls{client}s transfer sensitive data via the user's browser, which is deprecated and considered dangerous in \gls{sso}.
With respect to \gls{csrf}, we identified \num{447} logins with an entirely missing protection.
With \gls{spar}, we extended the list of vulnerabilities with \num{337} additional logins due to a recognized weak \gls{csrf} protection, for instance, less then 20 bytes entropy.
Furthermore, we identified nested URL parameters in \num{9} cases that could lead to open redirects and \num{5} \gls{client} secret leakages with \gls{spar}'s deep inspection.%

\paragraph{RQ3: How private are current \gls{sso} logins?}
To authenticate users, \glspl{client} and \glspl{idp} exchange sensitive user-related information.
This exchange must only happen transparently and after the user's consent.
We are the first to identify that this is not always the case.
We estimated \num{200} 
cases in which the \glspl{client} and \glspl{idp} %
exchange private user information secretly without user awareness.

\paragraph{\glsentrytext{ssomon}}
\gls{ssomon} is our answer to RQ1, RQ2, and RQ3.
It can conduct an automated evaluation of large-scale \gls{sso} deployments using the Tranco \num{10k}.
We depict its basic idea in \autoref{fig:intro}.
Once it detects \gls{sso} support on a website, \gls{ssomon} sequentially signs in using Apple, Facebook, and Google.
Next, \gls{ssomon} repeats the automated login a second time so that it can distinguish random from static parameters.
\gls{ssomon} then automatically identifies security and privacy issues.
\gls{ssomon} combines novel insights on how \gls{sso} schemes are implemented with state-of-the-art engineering techniques.

\paragraph{Contributions}
We make the following key contributions:
\begin{itemize}[label=$\blacktriangleright$, noitemsep]
    \item We systematize known \gls{sso} analysis techniques (\autoref{sec:sso_tools}).
    We compare prior tools in \autoref{tab:tools}, and we show how \gls{ssomon} differs to them with novel \gls{sso} detection and analysis techniques.
    
    \item We present \gls{ssomon}, our systematic and modular approach for large-scale \gls{sso} analyses (\autoref{sec:ssomon}).
    \gls{ssomon} is open-source\footnote{For the submission, we provide the source code and screenshots of \gls{ssomon} on \num{\url{https://tinyurl.com/sso-monitor}}. For anonymity reasons, we will publish the artifacts after the review phase.} and only requires a list of domains as input.
    It identifies the login page for each domain, detects which \glspl{idp} are supported, and starts the authentication process.
    \gls{ssomon} records the traffic that it later analyzes on security and privacy issues.
        
    \item We publish an in-depth overview of the usage of \gls{sso}
    in the Tranco \num{1k} (manual ground truth analysis $\rightarrow$ \autoref{sec:ground_truth})
    and Tranco \num{10k} (automated analysis with \gls{ssomon} $\rightarrow$ \autoref{sec:landscape})
    to answer RQ1.
    
    \item We use \gls{ssomon} to analyze the security of \num{3,020} \gls{sso} logins across \num{1,632} websites to answer RQ2 (\autoref{sec:security}).
    Besides \num{337} weak and \num{447} missing \gls{csrf} protections, it uncovers \num{342} obsolete protocol flows, and \num{215} protocol mix-ups.
    Since \gls{ssomon} inspected nested and encoded data structures, it identified \num{5} \gls{client} secret leakages, and \num{an} open redirect attack. %

    \item We reveal \num{200} privacy breaches among \num{1,632} websites that support \gls{sso} to answer RQ3 (\autoref{sec:privacy}).
    We are the first to identify that websites secretly sign in their users using \gls{sso} without their knowledge and awareness.
    
\end{itemize}

\paragraph{Responsible Disclosure}
We notified the affected sites as part of our ongoing responsible disclosure process to achieve a more secure and private \gls{sso} landscape.
We relied on well-established security reporting mechanisms from prior work~\cite{drakonakis2020cookie, stock2016, 197270, cispa1190} to collect the contact emails:
\begin{inparaenum}
    \item the \code{security.txt} file
    \item the WHOIS record
    \item off-the-shelf search engine~\cite{maldevel64:online} and website~\cite{arifszn2022Oct} email crawlers, and
    \item the standard aliases \code{security@}, \code{abuse@}, \code{webmaster@}, and \code{info@}.
\end{inparaenum}
We sent the email from our institutional email address to verify our identity and maximize credibility.
While we participate in the active discussions with the vendors, some of them have resolved the issues.
We appreciate for being acknowledged and rewarded with bug bounties.

%% file: sections/20_basics.tex
\section{Background: \glsentrylong{sso} Schemes}
\label{sec:basics}

\begin{figure}[t]
    \centering
    \includegraphics[width=\linewidth]{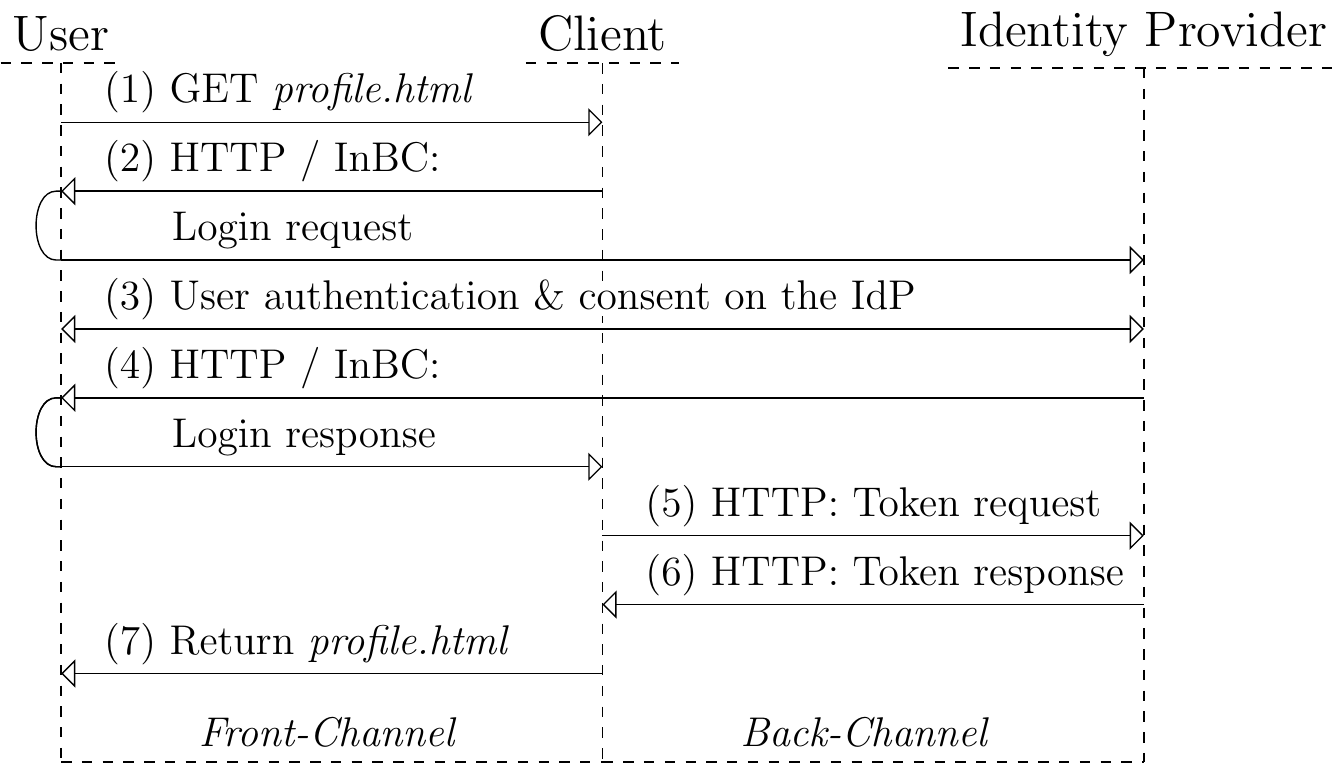}
    \caption{
    \gls{sso} Scheme.
    The \gls{client} website delegates the \gls{enduser}'s authentication to the \gls{idp}.
    The \gls{enduser}'s browser transfers messages in the \gls{fc} using \gls{http} or \glsentryshortpl{ibc} with \gls{js} (1-4, 7).
    \Gls{bc} messages (5-6) are based on \gls{http} and invisible in the \gls{enduser}'s browser.}
    \label{fig:oidc}
\end{figure}

\autoref{fig:oidc} depicts a basic \gls{sso} scheme.
It consists of an \gls{enduser} who wants to log in on the \gls{client}'s website using an \gls{idp}.
\Gls{sso} protocols can be divided into \gls{fc} communication, which is exposed to the \gls{enduser}'s browser (steps 1-4 and 7), and \gls{bc} communication (steps 5-6), which is invisible to the \gls{enduser}.
\Gls{sso} messages in the \gls{fc} can be sent via \gls{http} or \gls{ibc} techniques, such as the \gls{pm} \gls{api}~\cite[\S9.3]{whatwgHTMLLivingStandard} and \gls{cm}~\cite[\S9.4]{whatwgHTMLLivingStandard} \gls{api}.
\Gls{http} communication is standardized in \gls{sso}, but recent research~\cite{jannett2022} showed a strong shift towards the use of \gls{ibc} techniques in \gls{sso}, which rely on \gls{js}.

\paragraph{Login Request and Response}
The \gls{sso} login flow starts with the \gls{enduser} requesting access to a restricted resource, for example, to \emph{profile.html}.
To authenticate the \gls{enduser}, the \gls{client} sends the \gls{lreq} to the \gls{idp} via the \gls{enduser}'s browser.
This message contains parameters specific to the particular \gls{sso} protocol in use.
In \gls{oidc}, the \gls{lreq} contains the identity of the \gls{client} (\code{client_id}), the target to which the \gls{idp} must send the \gls{lres} (\code{redirect_uri}), and optional security parameters (i.e., \code{state}, \code{nonce}, \code{code_challenge}, \ldots).
The \gls{lres} contains the tokens (\code{code}, \code{access_token}, \code{id_token}) that the \gls{client} uses to authenticate the \gls{enduser}.

\paragraph{User Authentication \& Consent on the \gls{idp}}
Before the \gls{lres} can be sent back to the \gls{client}, the \gls{enduser} must authenticate to the \gls{idp}.
Protocols based on \gls{oauth}, such as \gls{oidc} or \gls{fbc}, also ask the \gls{enduser} to provide consent on resources to be accessed by the \gls{client}.

\paragraph{Token Request and Response}
\Gls{sso} protocols can authenticate the \gls{enduser} either by using only the information in the \gls{lres}, or by using the \gls{bc}.
\Gls{oidc} offers both variants, which can be configured in the \gls{lreq} using dedicated parameters (\gls{response_type}).
If the \gls{bc} authentication is used, the \gls{client} sends a \gls{treq} to the \gls{idp}.
This \gls{http} message contains authentication information of the \gls{client} (e.g., \code{client_id} and \code{client_secret}) as well as information from the \gls{lres}.
In \gls{oidc}, the \gls{lres} contains a \code{code}, a one-time-use token that is bound to the \gls{client}.
Once the \gls{client} redeems the \code{code} in the \gls{treq} on the \gls{idp}, it retrieves the \gls{tres} that holds a \gls{jwt} with the \gls{enduser}'s identity.

%% file: sections/25_sso_tools.tex
\section{Systematization of Known \glsentryshort{sso} Tools}
\label{sec:sso_tools}

We investigated related work on tools performing \gls{sso} security and privacy analyses or using \gls{sso} for automated account sign-ins and registrations.
\autoref{tab:tools} summarizes our comparison grouped into categories that we found useful for answering RQ1-3.

\input{sections/25_table_sso_tools.tex}

\paragraph{Availability}
The minority of \gls{sso} analysis tools (\num{6}/\num{19}) are publicly available.
We strongly believe that this attitude prevents the community from proceeding with research and yields the reimplementation of already solved tasks.
For instance, all related works implement individual \gls{sso} automation pipelines instead of reusing existing ones.
Thereby, we provide \gls{ssomon} as an open-source foundation for future large-scale analyses of the Internet's \gls{sso} ecosystem.

\paragraph{\glsentryshort{sso} Scope}
To execute \gls{sso}, the authentication and consent steps must be automated for \emph{each} \gls{idp}.
For this reason, prior tools often support a \emph{single} \gls{idp} (\num{5}/\num{19}). %
With \gls{ssomon}, we support the three most popular \glspl{idp}~\cite{morkonda2021}: Apple, Facebook, and Google. 
Interestingly, prior tools work best for English websites due to their pattern-based \gls{sso} detection.
This is mirrored in the selection of top sites lists, for example, \citet{ssoscan} use the region-specific Alexa list.
\gls{ssomon}'s visual-based \gls{sso} detection works for websites of all languages and regions.
We also consider \glspl{ibc} on a large scale, which recently got attention in \gls{sso}~\cite{jannett2022}.

\paragraph{Login Page Detection}
Prior work used different approaches to find the login page
For example, they tested for common paths (i.e., \url{/login})~\cite{jonker2018shepherd,ghasemisharif2022},
only crawling links including login-related keywords ($\rightarrow$ selective),
or running a \gls{dfs} crawl~\cite{Shernan2015,authscope,ghasemisharifSingleSignOffWhere2018,jonker2018shepherd,drakonakis2020cookie,morkonda2021}.
Others~\cite{ssoscan} assumed the homepage as login page~\cite{ssoscan,ghasemisharifSingleSignOffWhere2018,jonker2018shepherd,drakonakis2020cookie}.
In practice, websites can include the \gls{sso} button on a deeply nested login page.
Thus, starting with the input of a domain, tools first have to aggregate a candidate pool of login pages.
Therefore, we assess two techniques:
\begin{inparaenum}
    \item \gls{bfs} crawling with a depth of 2 visiting links including login-related keywords first ($\rightarrow$ prioritized), and
    \item querying search engines.
\end{inparaenum}

\paragraph{\glsentryshort{sso} Detection}
The login page candidates are scanned for \gls{sso} buttons.
All prior work followed a programmatic, \emph{pattern-based} detection approach, which uses string-matching algorithms to identify keywords.
For instance, if a \html{<button>} contains the keyword \emph{login}, it is tested for \gls{sso}.
This approach suffers from \glspl{fn}, as \gls{sso} buttons can take any shape and include arbitrary keywords in any language.
For example, they can be \html{<button>} tags with \code{onclick} listeners or nested like \html{<a><img /></a>}.
\gls{ssomon} introduces a novel, \emph{visual-based} detection approach, which identifies the \glspl{idp}' logos contained in \gls{sso} buttons.
We randomly sampled a subset of \num{50} websites with \gls{sso} and found that \num{49} of them include logos in all of their \gls{sso} buttons.
\gls{ssomon} combines the keyword-based approach with the novel visual-based approach to maximize the detection rate on any website.

\paragraph{\glsentryshort{sso} Execution}
We found that \num{5} of \num{19} tools could execute \gls{sso} to login on the \gls{c}.
However, SSOScan~\cite{ssoscan} was last updated in 2015 and not adapted to today's \gls{sso} logins.
AuthScope~\cite{authscope} is for mobile apps.
Shepherd~\cite{jonker2018shepherd} and Cookie Hunter~\cite{drakonakis2020cookie} both use \gls{sso} logins as fallback for generic post-authentication studies.
\gls{ssomon} is the first to automate the Apple login, including its \gls{2fa}.

\paragraph{Account Registration and Verification}
Studies focused on post-authentication mechanisms also require post-\gls{sso} account registration and verification.
For instance, the user is asked to submit additional data that is not provided by the \gls{idp} and confirm the email address.
Therefore, email verification~\cite{drakonakis2020cookie,ghasemisharif2022}, \gls{sms} verification~\cite{ghasemisharif2022}, and \glspl{captcha}~\cite{ghasemisharif2022} have been automated.
For \gls{ssomon}, we consider them as out of scope as we examine the protocol messages that are \emph{always} exchanged during \emph{any} login.
Also, \gls{ssomon} is open source and releasing automated account registration tools to the public may raise ethical concerns.

\paragraph{Continuous Monitoring}
To the best of our knowledge, \gls{ssomon} is the first to provide a constantly updated top sites list of websites with \gls{sso}, similar to Tranco~\cite{tranco}.
Only SAAT~\cite{ghasemisharif2022} recently compared the \gls{sso} landscape over a period of 50 days.
They noticed a dynamic landscape, which \gls{ssomon} is the first to monitor continuously.

\paragraph{Ground Truth Estimation}
We compare the automated \gls{sso} detection engine of \gls{ssomon} against the Tranco \num{1k} to estimate its accuracy.
Prior work~\cite{ssoscan, jonker2018shepherd, drakonakis2020cookie} randomly sampled only small subsets of websites for such estimations (i.e., \num{20}~\cite{drakonakis2020cookie}, \num{50}~\cite{jonker2018shepherd}, and \num{169}~\cite{ssoscan}).
In sum, \gls{ssomon} detects \num{97\%} of all \gls{sso} login buttons, and it executes a total of \num{2,811} \gls{sso} logins.

\paragraph{\glsentryshort{sso} Analyses}
\gls{ssomon} runs a comprehensive and systematic study on the real-world adoption of the \gls{oauth} security best practices~\cite{lodderstedtOAuthSecurityBest}.
Prior studies~\cite{ssoscan, Shernan2015, yang2016model, yang2018, li2019oauthguard, wei2021} already investigated selected parameters (i.e., \code{state}) but missed to conduct in-depth parameter inspections.
With \gls{spar}, we fill this gap.
Regarding privacy, \gls{ssomon} first reveals that websites are logging in to their \glspl{enduser} secretly without their awareness.

%% file: sections/25_table_sso_tools.tex
\newcommand{\osTnote}{1}
\newcommand{\listsTnote}{2}
\newcommand{\lpsourceTnote}{3}
\newcommand{\crawlTnote}{4}
\newcommand{\enginesTnote}{5}
\newcommand{\recmethTnote}{6}
\newcommand{\fnTnote}{7}

\newcommand{\tnote}[1]{$^{#1}$}
\newcommand{\Shortunderstack}[1]{#1}

\newcommand{\scopeWeb}{\faGlobe}
\newcommand{\scopeMobile}{\faMobile}
\newcommand{\loginsManual}{--}

\newcommand{\tool}[3]{#1~#2 (#3)}

\begin{table*}[t]
\resizebox{\textwidth}{!}{%
\def\arraystretch{1.25}
\begin{tabular}{lc|ccccc|cccccc|cc|cc}
\toprule
\multicolumn{2}{c|}{\textbf{Availability}} & \multicolumn{5}{c|}{\textbf{\gls{sso} Scope}} & \multicolumn{8}{c|}{\textbf{\gls{sso} Detection \& Execution}} & \multicolumn{2}{c}{\textbf{\gls{sso} Analyses}} \\
Tool \& Publication & OS\tnote{\osTnote} & \glspl{client}\tnote{\listsTnote} & \glspl{idp} & \shortstack{Web /\\Mobile} & \shortstack{\glsentryshort{oauth} /\\\glsentryshort{oidc}?} & \glsentryshort{ibc}? & \shortstack{Login Page\tnote{\lpsourceTnote,\crawlTnote,\enginesTnote}\\Detection} & \shortstack{\gls{sso}\tnote{\recmethTnote}\\Det.} & \shortstack{\gls{sso}\\Exe.} & \shortstack{Acc.\tnote{\recmethTnote}\\Reg.} & \shortstack{Acc.\\Verif.} & Monitor & \shortstack{\# \gls{sso}\\Logins} & \shortstack{False\\Negatives} & Security & Privacy \\
\midrule
\tool{BRM Analyzer}{\cite{Wang2012}}{S\&P'12} & \nsup & 11 & \faGoogle~\faFacebook~\faPaypal~+1 & \scopeWeb & \nsup & \nsup & \nsup & \nsup & \nsup & \nsup & \nsup & \nsup & -- & -- & \fsup & \nsup \\
\tool{AuthScan}{\cite{authscan}}{NDSS'13} & \nsup & 8 & \faFacebook~+2 & \scopeWeb & \nsup & \nsup & \nsup & \nsup & \nsup & \nsup & \nsup & \nsup & -- & -- & \fsup & \nsup \\
\tool{SSOScan}{\cite{ssoscan}}{USENIX'14} & \href{https://github.com/Treeeater/vulCheckerFirefox}{\fsup} & QC~20k & \faFacebook & \scopeWeb & \fsup & \nsup & HP-Prio & \faCode & \fsup & \fsup & \nsup & \nsup & 1,660 & 3\% (5/169) & \fsup & \nsup \\
\tool{OAuth Detector}{\cite{Shernan2015}}{DIMVA'15} & \href{https://bitbucket.org/uf_sensei/oadpublic/src/master/}{\fsup} & AX~10k & Any \glsentryshort{idp} & \scopeWeb & \fsup & \nsup & Crawl-DFS-D2 & \faCode & \nsup & \nsup & \nsup & \nsup & 302 & -- & \fsup & \nsup \\
\tool{MPWA-ZAP}{\cite{blackboxMPWA_NDSS16}}{NDSS'16} & \nsup & -- & \faFacebook~\faLinkedin~\faPaypal~\faInstagram & \scopeWeb & \fsup & \nsup & \nsup & \nsup & \nsup & \nsup & \nsup & \nsup & -- & -- & \fsup & \nsup \\ 
\tool{OAuthTester}{\cite{yang2016model}}{AsiaCCS'16} & \nsup & AX~500 & \faFacebook~+3 & \scopeWeb & \fsup & \nsup & \nsup & \nsup & \nsup & \nsup & \nsup & \nsup & 405 & \loginsManual & \fsup & \nsup \\
\tool{PrOfESSOS}{\cite{sok_oidc}}{EuroS\&P'17} & \href{https://github.com/RUB-NDS/PrOfESSOS}{\fsup} & 8 libraries & -- & \scopeWeb & \fsup & \nsup & \nsup & \nsup & \nsup & \nsup & \nsup & \nsup & -- & -- & \fsup & \nsup \\
\tool{AuthScope}{\cite{authscope}}{CCS'17} & \nsup & GP~200k & \faFacebook & \scopeMobile & \nsup & \nsup & Crawl-DFS-D$\infty$-Prio & \faCode & \fsup & \nsup & \nsup & \nsup & 4,838 & -- & \nsup & \nsup \\
\tool{Single Sign-Off}{\cite{ghasemisharifSingleSignOffWhere2018}}{USENIX'18} & \nsup & AX~1m & \faGoogle~\faFacebook~\faTwitter~\faLinkedin~+61 & \scopeWeb & \fsup & \nsup & \Shortunderstack{HP; Crawl-BFS-D2-Select;\\CPs; SEs-DDG} & \faCode & \nsup & \nsup & \nsup & \nsup & 57,555 & -- & \fsup & \nsup \\
\tool{S3kvetter}{\cite{yang2018}}{USENIX'18} & \nsup & -- & 10~\glsentryshortpl{sdk} & \scopeWeb & \fsup & \nsup & \nsup & \nsup & \nsup & \nsup & \nsup & \nsup & -- & -- & \fsup & \nsup \\
\tool{OAUTHLINT}{\cite{rahat2019}}{ASE'19} & \nsup & GP~600 & \faGoogle~\faFacebook~\faTwitter~+17 & \scopeMobile & \fsup & \nsup & \nsup & \faCode & \nsup & \nsup & \nsup & \nsup & 316 & -- & \fsup & \nsup \\
\tool{OAuthGuard}{\cite{li2019oauthguard}}{SSR'19} & \href{https://github.com/oauthguard/OAuthGuard}{\fsup} & MJ~1k & \faGoogle & \scopeWeb & \fsup & \nsup & \nsup & \nsup & \nsup & \nsup & \nsup & \nsup & 137 & \loginsManual & \fsup & \fsup \\
\tool{Shepherd}{\cite{jonker2018shepherd}}{MADWeb'20} & \nsup & AX~10k & \faFacebook & \scopeWeb & \fsup & \nsup & \Shortunderstack{HP; Crawl-D2-Select;\\CPs; SEs-SP,BI,ASK} & \faCode & \fsup & \nsup & \nsup & \nsup & 664 & 18\% (9/50) & \nsup & \nsup \\ 
\tool{Cookie Hunter}{\cite{drakonakis2020cookie}}{CCS'20} & \nsup & AX~1.5m & \faGoogle~\faFacebook & \scopeWeb & \fsup & \nsup & \Shortunderstack{\cite{ghasemisharifSingleSignOffWhere2018}; Crawl-BFS-D2-Prio;\\HP-Top30} & \faCode & \fsup & \fsup & Email & \nsup & 2,066 & 25\% (5/20) & \nsup & \nsup \\ 
\tool{OAuthScope}{\cite{morkonda2021}}{WPES'21} & \nsup & AX~500*5 & \faGoogle~\faFacebook~\faApple~\faLinkedin & \scopeWeb & \fsup & \nsup & Crawl-Select & \faCode & \nsup & \nsup & \nsup & \nsup & 815 & \loginsManual & \nsup & \fsup \\
\tool{MoScan}{\cite{wei2021}}{ISSTA'21} & \href{https://github.com/baigd/moscan}{\fsup} & MZ~26 & \faFacebook~\faTwitter~\faLinkedin~\faGithub & \scopeWeb & \fsup & \nsup & \nsup & \nsup & \nsup & \nsup & \nsup & \nsup & 26 & \loginsManual & \fsup & \nsup \\
\tool{SAAT}{\cite{ghasemisharif2022}}{S\&P'22} & \nsup & MJ~100k & \faFacebook & \scopeWeb & \fsup & \nsup & SEs-SP,BI,DDG; CPs & \faCode & \fsup & \fsup & \Shortunderstack{Email, SMS, \\ CAPTCHA} & 50 days & 2,689 & -- & \fsup & \nsup \\ 
\tool{Cerberus}{\cite{rahat2022}}{CCS'22} & \nsup & -- & 10~libraries & \faGlobe & \fsup & \nsup & \nsup & \nsup & \nsup & \nsup & \nsup & \nsup & -- & -- & \fsup & \nsup \\ 
\tool{DISTINCT}{\cite{jannett2022}}{CCS'22} & \href{https://github.com/RUB-NDS/DISTINCT}{\fsup} & TC~1k & \faGoogle~\faFacebook~\faApple & \scopeWeb & \fsup & \fsup & \nsup & \nsup & \nsup & \nsup & \nsup & \nsup & 273 & \loginsManual & \fsup & \fsup \\
\midrule
\midrule
SSO-Monitor & \fsup & TC~10k & \faGoogle~\faFacebook~\faApple & \scopeWeb & \fsup & \fsup & \Shortunderstack{Crawl-BFS-D2-Prio; SEs-SP,BI,DDG} & \Shortunderstack{\faEye + \faCode} & \fsup & \nsup & \nsup & \Shortunderstack{Continuous} & 2,953 & 3\% (in 1k) & \fsup & \fsup \\ \bottomrule
    \multicolumn{17}{l}{\osTnote~\textbf{Availability} --- OS: Open Source; } \\
    \multicolumn{17}{l}{\listsTnote~\textbf{Top Sites / Apps Lists} --- AX: Alexa; TC: Tranco; MJ: Majestic; MZ: Moz; QC: Quantcast; GP: Google Play;} \\
    \multicolumn{17}{l}{\lpsourceTnote~\textbf{Login Page Sources} --- HP: Homepage; Crawl: Crawling; CPs: Common Paths; SEs: Search Engines;} \\
    \multicolumn{17}{l}{\crawlTnote~\textbf{Crawling Techniques} --- \glsentryshort{dfs}: \glsentrylong{dfs}; \glsentryshort{bfs}: \glsentrylong{bfs}; Prio: Prioritized; Select: Selective; D: Depth;} \\
    \multicolumn{17}{l}{\enginesTnote~\textbf{Search Engines} --- SP: Startpage; BI: Bing; DDG: DuckDuckGo; ASK: Ask.com;} \\
    \multicolumn{17}{l}{\recmethTnote~\textbf{Recognition Methods} --- \faCode: Pattern- and Heuristics-based Recognition; \faEye: Visual-based Recognition;} \\
\end{tabular}
}
\caption{
Systematization of Known \glsentryshort{sso} Analysis Tools.
With the release of \gls{ssomon}, we contribute to the scope, detection, execution, and security and privacy analyses of \gls{sso} in the wild.
}
\label{tab:tools}
\end{table*}

%% file: sections/30_sso_monitor.tex
\section{\glssc{ssomon}: Design and Architecture}
\label{sec:ssomon}

\begin{figure}[t]
    \centering
    \includegraphics[width=\linewidth]{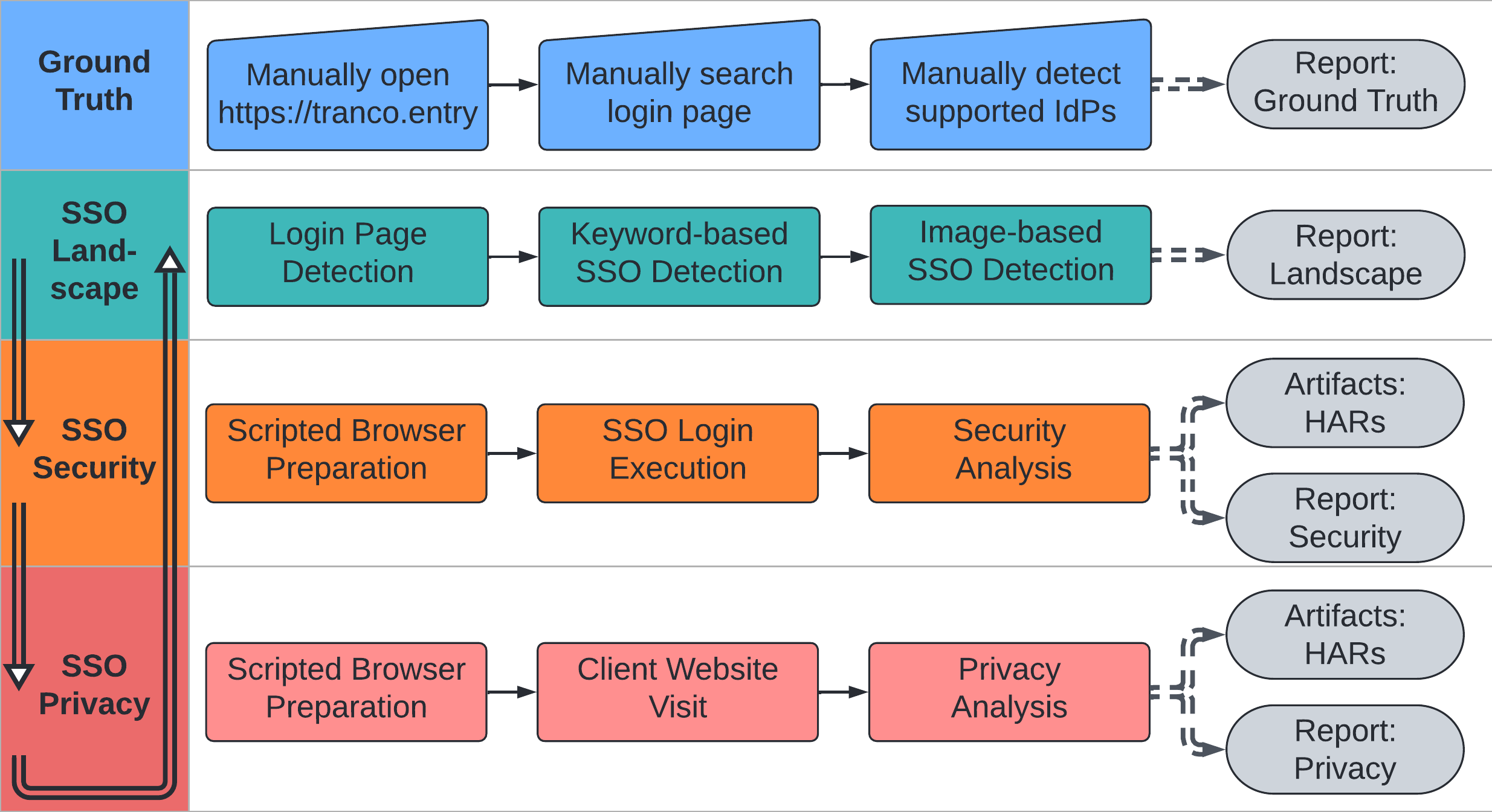}
    \caption{
    Design and Architecture of \gls{ssomon}.
    It fully automates the \gls{sso} analysis in four steps:
    (1) ground truth estimation,
    (2) landscape evaluation,
    (3) security analysis, and
    (4) privacy analysis.
    Only the first step requires user interaction
    and is executed \emph{once}
    to estimate its detection accuracy.
    }
    \label{fig:sso_monitor}
\end{figure}

In this section, we introduce the design and the architecture of \gls{ssomon}, see \autoref{fig:sso_monitor}.
On input of a domain, \gls{ssomon} finds the login page, the supported \glspl{idp}, executes the \gls{sso} login, and derives security and privacy results.

\paragraph{Design}
\Gls{ssomon} can investigate the \gls{sso} landscape, its security, and its privacy fully automatically.
\autoref{fig:sso_monitor} depicts its general idea that is split into four modules.
The \emph{ground truth} is our initial manual investigation of the \gls{sso} landscape, see \autoref{sec:ground_truth}.
\gls{ssomon} guides the analyst via an interactive interface to configure the \gls{sso} support for a specific website.
By contrast, the \emph{landscape detection} works fully automatically, see \autoref{sec:landscape}.
To estimate the automatic detection's success rate, we compare our results with the ground truth.
The remaining two parts, \emph{\gls{sso} security} (\autoref{sec:security}) and \emph{\gls{sso} privacy} (\autoref{sec:privacy}) conduct multiple automatic sign-ins based on the automatic landscape detection.
Both record the \gls{http} traffic and all \glspl{ibc} during these sign-ins for their actual analysis.

\paragraph{Architecture}
The application architecture of \gls{ssomon} consists of a master node and a variable number of worker nodes.
The master distributes all analysis tasks to the workers on a per domain basis.
Therefore, \gls{ssomon} scales efficiently by adding new workers.
All artifacts and reports are centrally collected and stored on the master node.
Additionally, the master provides a web-based management interface.
Hence, all administrative tasks and analysis reports can easily be executed and viewed.

%% file: sections/40_ground_truth.tex
\section{Ground Truth: Tranco 1k \glsentryshort{sso} Landscape}
\label{sec:ground_truth}

We need a ground truth to implement and evaluate our automated discovery and analysis of \gls{sso}.
We manually analyzed the websites out of the ground truth, and we use our analysis results as reference for \gls{ssomon}'s automatic evaluations.
In our case, the ground truth allows us to choose proper strategies for the automation of \gls{sso}.
This process includes choosing appropriate parameters and fine-tuning our automated \gls{sso} detection engine for high-level accuracy.

\paragraph{Methodology}
We use the Tranco list\footnote{Available at https://tranco-list.eu/list/6Z2X.}~\cite{tranco} generated on 15 November 2021 as the foundation for our manual investigations.
For each domain in the list, we manually visited the appropriate website.
We looked for login possibilities on the site, and if available, documented the \emph{supported \gls{sso} providers} that can be used to sign in.
We further noted the \emph{domain}, \emph{login page}, \emph{timestamp}, and \emph{automation hurdles}. %

\paragraph{Results}
In \autoref{tab:ground_truth}, we depict the results of our ground truth analysis.
Out of the Tranco \num{1k}, \num{760} websites (\num{76}\%) implement a login page and thus support user authentication.
The remaining sites do not implement user authentication (\num{106, ~11\%}) or are not reachable (\num{134, ~13\%}).
We further found that \num{278} sites (\num{28\%)} support \gls{sso} with at least one \gls{idp}. 
The most used \gls{idp} is Google on \num{244} websites (\num{24\%}), strictly followed by Facebook on \num{214} sites (\num{21\%}).
Apple started its \gls{sso} support back in 2019.
Since then, its importance has significantly grown, as it is already supported on \num{122} websites (\num{12\%}).
We determined that the automated detection and execution of \gls{sso} is not possible on \num{55} sites due to technical constraints, such as required user interaction.
Thus, we evaluate the success rate of our automated approach against \num{223} websites and \num{463} \gls{sso} logins, respectively.

\begin{table}[t]
\centering
\resizebox{\columnwidth}{!}{%
\begin{tabular}{lc|cccc}
    \toprule
    & \textbf{Websites} & \faGoogle & \faFacebook & \faApple & $\sum$ \textbf{Logins} \\
    \midrule
    Authentication supported        & 760 (76\%) & -- & -- & -- & -- \\
    \gls{sso} Login supported       & 278 (28\%) & 244 & 214 & 122 & 580 \\
    Automated \gls{sso} possible    & 223 (22\%) & 197 & 167 & 99  & 463 \\
    \bottomrule
\end{tabular}
}
\caption{Ground Truth. \gls{sso} is supported on \num{28}\% of the Tranco \num{1k}. On \num{22}\%, we can execute \gls{sso} fully automated.}
\label{tab:ground_truth}
\end{table}

%% file: sections/50_landscape.tex
\section{Automatic Evaluation: \glsentryshort{sso} Landscape}
\label{sec:landscape}

In this section, we present our \gls{sso} landscape evaluation, which contains the
methodology (\autoref{sec:landscape_methodology}),
login page discovery (\autoref{sec:crawling}, \autoref{sec:search_engines}),
\gls{sso} detection (\autoref{sec:keyword_recognition}, \autoref{sec:logo_recognition}),
detection rate and performance (\autoref{sec:verification_recognition}),
continuous monitoring (\autoref{sec:continous_monitoring}),
and contemporary Tranco \num{10k} \gls{sso} landscape (\autoref{sec:landscape_results}).

\subsection{Methodology}
\label{sec:landscape_methodology}
In this paper, we concentrate on the continuous monitoring of \gls{sso} in the wild.
The monitoring requires computational resources over a long period.
Thus, we define the following requirements for the detection:
\begin{inparaenum}
\item as few resources as possible as much as needed, 
\item high detection rate, and
\item robustness.
\end{inparaenum}

\paragraph{Available Resources}
Our goal is to monitor a large number of websites once a year continuously.
Thus, the required resources to discover the login page and analyze the \gls{sso} support should not exceed 52 minutes in the worst case:
\begin{equation}
max_{time} = \frac{12(month) * 30(d)* 24 (h) * 60 (m)}{10.000 (websites)* 1(worker)} = 52~min/site  
\end{equation}

We also require high detection rates with low \glspl{fp} and \glspl{fn}.
Our goal is to achieve an at least 90\% correct detection of \gls{sso}.
Many of the websites, however, are not available in English.
We should be able to analyze and discover \gls{sso} on such websites. 

\paragraph{Preparation Steps}
\label{sec:preparation_recognition}
\begin{inparaenum}
    \item We use Selenium to automate the navigation on websites.
    However, many websites detect when the browser is automatically navigated and activate \glspl{captcha}.
    To circumvent this limitation, we use the \texttt{selenium-stealth}\footnote{\url{https://pypi.org/project/selenium-stealth/}} plugin.
    \item We disable the cookie banners by installing the browser extension \texttt{I don't care about cookies}\footnote{\url{https://addons.mozilla.org/en-US/firefox/addon/i-dont-care-about-cookies/}} and thus reduce the risks of manual interactions to a minimum.
    \item To execute Selenium on headless servers, we use a virtual monitor~\footnote{\url{https://pypi.org/project/virtualenv/}} and 
    the Google Chrome browser with a pre-configured window size.
\end{inparaenum}

\subsection{Login Page Discovery: Crawling}
\label{sec:crawling}

Crawling is used by the majority of related work~\cite{authscope, ghasemisharifSingleSignOffWhere2018, Shernan2015, jonker2018shepherd, drakonakis2020cookie} to discover login pages.
However, we suggest that this approach suffers from low reliability and cost-benefit ratio.

\paragraph{Methodology}
We configured Scrapy\footnote{\url{https://scrapy.org/}} to \gls{bfs}-crawl the Tranco \num{1k} from our ground truth with a depth of 2.
We took advantage of Scrapy's built-in \code{LinkExtractor} module, which automatically detects and extracts all clickable links on a page.
To eliminate \glspl{fp}, we filtered out third-party links. %
We followed crawling best practices~\cite[\S5]{182948} like the \code{robots.txt} denylist and adaptive request throttling.

\paragraph{Results}
Our crawling dataset contains \num{515,855} links from \num{1k} websites, averaging to \num{515} crawled links per site.
However, the entire crawling set contains only \num{146} login pages out of the ground truth (\num{760}).
Crawling statistics show that \num{104 (71\%)} of the login pages are linked on the homepage, while \num{42 (29\%)} of them are linked on a subpage.
The search engine approach, see \autoref{sec:search_engines}, detects almost three times more login pages.
On top, continuous monitoring requires the crawling to be run on a regular basis.
This puts excessive load onto the web servers.
Due to the ambitious resource load, the approach does not scale and does not satisfy our research goal.

\subsection{Login Page Discovery: Search Engines}
\label{sec:search_engines}

Search engines provide benefits out-of-the-box:
\begin{inparaenum}
    \item they already crawl the web with an indefinite depth,
    \item they instantly provide up-to-date results,
    \item they make the data accessible and searchable via keywords, and
    \item they use internal rankings to provide optimized results.
\end{inparaenum}
Prior \gls{sso} tools~\cite{ghasemisharifSingleSignOffWhere2018, jonker2018shepherd, ghasemisharif2022} used search engines but did not systematically evaluate their effectiveness.
We answer the following questions:
\begin{inparaenum}
    \item Which search engines return the most login pages?
    \item Which search query returns the most login pages?
    \item How many ranked search results are required to detect the most login pages?
\end{inparaenum}

\paragraph{Methodology}
According to~\cite{searchengines}, Google (\num{92.01\%}), Bing (\num{2.96\%}), Yahoo (\num{1.51\%}), Baidu (\num{1.17\%}), YANDEX (\num{1.06\%}), and DuckDuckGo (\num{0.68\%}) are the most popular search engines in 2022.
Yahoo uses Bing's search index.
Baidu and YANDEX primarily target the Chinese or Russian market.
Thus, we included Google, Bing, and DuckDuckGo in our scope.
We further replaced Google with Startpage, which proxies the Google Search and does not interfere with \glspl{captcha}.
To eliminate \glspl{fp}, we require the website and its login page to be the same site using the \code{site:} operator.
For each resolved domain, we submitted five different search queries to each engine and stored the top 10 returned search results.
We selected appropriate search queries to the best of our knowledge ranging from simple to more specific ones:
{\small
\begin{inparaenum}
    \item \texttt{reddit.com login site:reddit.com}
    \item \texttt{reddit login site:reddit.com}
    \item \texttt{log in reddit site:*.reddit.com}
    \item \texttt{reddit login signin signup register account site:reddit.com}
    \item \texttt{site:reddit.com (intitle:"login" OR intitle:"log in" OR intitle:"signin" OR intitle:"sign in")}
\end{inparaenum}
}

\paragraph{Results}
Surprisingly, the most straightforward search query ($SQ1$) combined with all engines found the most login pages out of the ground truth (\num{434/760}), see \autoref{tab:search_engines_eval}.
Although we can compare different engines and queries \emph{with each other}, this number only indicates a \emph{lower boundary}.
The ground truth only holds a single login page for each website, while in practice, a website can have multiple login pages.
Our goal is to provide continuous monitoring of the \gls{sso} landscape once a year on a single-threaded machine (cf. \autoref{sec:continous_monitoring}).
Thus, performance plays an important role.
Our analysis takes \num{289} seconds for each page (cf. \autoref{tab:sso_recognition}).
The analysis of the top 3 search results from all three engines ($\rightarrow$ 9 in total) would require $\frac{289s \cdot 9 \cdot 10,000}{60 \cdot 60 \cdot 24} = 301~\text{days}$.
By using at least 4 parallel workers, we can scale up the landscape evaluation to a quarterly executed scan with a duration of 75 days.

\begin{table}[t]
\centering
\resizebox{\columnwidth}{!}{%
\begin{tabular}{lccccc}
    \toprule
    Login Pages found with  & $SQ1$     & $SQ2$     & $SQ3$     & $SQ4$     & $SQ5$     \\
    \midrule
    Startpage               & 307       & 329       & 250       & 216       & 150       \\ 
    Bing                    & 331       & 316       & 288       & 272       & 120       \\
    DuckDuckGo              & 320       & 315       & 264       & 268       & 132       \\
    Startpage \& Bing       & 423       & 414       & 370       & 343       & 172       \\
    Startpage \& DuckDuckGo & 420       & 416       & 363       & 342       & 179       \\
    Bing \& DuckDuckGo      & 354       & 340       & 310       & 306       & 144       \\
    All Combined (Top 10)   & 434       & 431       & 382       & 363       & 182       \\
    \midrule
    Top 5 Search Results    & 385 (-49) & 386 (-45) & 357 (-25) & 333 (-30) & 163 (-19) \\
    Top 3 Search Results    & 357 (-77) & 337 (-94) & 335 (-47) & 292 (-71) & 153 (-29) \\
    \bottomrule
\end{tabular}
}
\caption{
Search Query Evaluation.
To find a suitable search query and engine, we compare the detection rate of five queries on three engines against our ground truth.
}
\label{tab:search_engines_eval}
\end{table}

\subsection{\glsentryshort{sso} Detection: Patterns and Keywords}
\label{sec:keyword_recognition}
Inspired by previous work~\cite{ssoscan}, we decided to implement a keyword-based analysis.
We base our analysis on the extraction of \emph{clickable} elements on a website like links and buttons, but also elements with \gls{js} events.
To reduce the set of candidates starting \gls{sso}, we search for specific keywords inside the elements' texts and attributes (e.g., \emph{sign in with google}, \emph{login with facebook}).
If this does not return valid results, we search for elements including specific \gls{idp} names (e.g., \emph{google}, \emph{facebook}, \emph{apple}).
We store all candidates in a list to evaluate them later.
The limitations of the keyword-based analysis are:
\begin{inparaenum}
 \item The \gls{sso} detection is limited to the defined keywords. Different languages or source code without any of the keywords lead to \glspl{fn}.
 \item \gls{sso} buttons containing only the \gls{idp} logo without any text lead to \glspl{fn}.
\end{inparaenum}

\subsection{\glsentryshort{sso} Detection: Images and Logos}
\label{sec:logo_recognition}
To solve the limitations of the keyword-based approach, we developed an image-based analysis.
During our investigations, we discovered that \gls{sso} buttons also contain the logo of the corresponding \gls{idp}.
Thus, we implemented an algorithm for opening a login page candidate, taking a screenshot, highlighting recognized \gls{idp} logos, and extracting their coordinates.
For the logo recognition, we used a state-of-the-art algorithm that supports pattern matching, is available in Python, and is open source.
Thus, we chose the OpenCV algorithm\footnote{\url{https://docs.opencv.org/3.4/d4/dc6/tutorial_py_template_matching.html}}.
In the implementation, we solved the following challenges: logo collection, logo scalability, robustness, and high \gls{fp} rate.

\paragraph{Logo Collection}
To establish a set of \gls{idp} logos, we carefully analyzed \num{100} websites and extracted the commonly used logos.
In sum, we stored two or three logos for each \gls{idp}.
This list can be easily extended by storing new logos.

\paragraph{Logo Scalability}
One of the main challenges is the variable size of the logos on the websites since the pattern matching algorithm works only if the sizes are similar.
A suitable solution is to scale the website's screenshot with different factors and analyze it repeatedly for each scale factor.
We discovered that it is far more efficient to scale the logo instead of the screenshot during our implementation. 
This approach requires fewer resources, for instance, time, memory, and CPU.

\paragraph{Robustness}
The main challenge is to balance a high recognition rate with performance.
This requires us to adjust multiple parameters: the number of pre-configured logos, different logo scales, and detection scores.
Based on experiments and optimizations, we reduced the logo set to only used ones.
We also reduced the number of scale iterations to a minimum without sacrificing pattern matching results.
Since the OpenCV algorithm outputs a value with a matching score, we implemented upper and lower bounds to determine whether \gls{sso} is detected.
The algorithm stops if a match is above the upper bound due to the high matching confidence.

\paragraph{High \glsentrylong{fp} Rate}
On websites without \gls{sso}, the algorithm matches areas looking similar to the logos, i.e., \emph{G} for Google, \emph{O} for Apple, and interestingly \emph{t} for Facebook.
We eliminate all \glspl{fp} with a generic approach, see \autoref{sec:verification_recognition}.

\subsection{Detection Rates and Performance}
\label{sec:verification_recognition}
The keyword-based and image-based analyses detect suitable \gls{sso} candidates.
Still, their \gls{fp} rates are high.
To solve this problem, we designed a reliable and robust verification.

\paragraph{Detection Verification}
For each candidate, we store the coordinates on the login page.
During their verification, we automatically navigate the browser to these coordinates and click on the area.
If the browser sends the \gls{lreq} to an \gls{idp}, we know that \gls{sso} with the corresponding \gls{idp} is supported.
This verification eliminates the \gls{fp} rate.
Storing the coordinates leads to an unexpected advantage regarding the automated execution of \gls{sso}.
Websites are changing their source code, including the \gls{html} elements, quite often.
These changes make the automated execution via Selenium impossible since it navigates via element IDs or HTML trees.
However, the position of buttons is constant.
With the stored coordinates, we can reliably and repeatedly execute \gls{sso}.

\paragraph{Hybrid Approach}
We decided to chain both, the keyword-based and image-based \gls{sso} detection into a \emph{hybrid} approach.
First, we execute the less resource-intensive keyword-based analysis.
If it is not successful, the more resource-intensive image-based approach is triggered.

\begin{table}[t]
\centering
\resizebox{\columnwidth}{!}{%
\begin{tabular}{p{1.4cm}|c|cccc|c}
    \toprule
    \textbf{\multirow{2}{1.6cm}{Analysis Approach}}   & \textbf{\multirow{2}{1cm}{\gls{sso} Logins}} & \multicolumn{4}{c|}{\textbf{Detection Rates}} & \textbf{Duration} \\
                      &     & FPs & FNs & Recognized & Rate & $t_{avg}$ \\
    \midrule
    Keyword           & 463 & 0 & 20  & 443 & 95\%  & 1:26m\\
    Image             & 463 & 0 & 52  & 411 & 89\%  & 5:16m\\
    Hybrid            & 463 & 0 & 12  & 451  & 97\% & 4:49m \\
    \bottomrule
\end{tabular}
}
\caption{By combining the keyword-based and the image-based \gls{sso} recognition, we achieve a 97\% detection rate.}
\label{tab:sso_recognition}
\end{table}

\paragraph{Results}
In \autoref{tab:sso_recognition}, we analyze the Tranco \num{1k} sites that have a login page and do not require additional steps to start the \gls{sso}. %
From our ground truth, we expect to recognize 463 \gls{sso} logins.
The keyword-based approach recognizes 95\% of the \gls{sso} logins.
The image-based approach recognizes 89\% of the \gls{sso} logins.
Areas on the website looking similar to the logos are the main reason for the lower recognition rate.
The problem can be solved by marking multiple candidates where the logo \emph{could be}.
Considering the resource and time restraints, we decided to address this improvement in future work.
By combining both approaches, we achieve a recognition rate of 97\%.
Interestingly, \num{6} of \num{12} \glspl{fn} are caused by \glspl{captcha} on \emph{ebay.com} and \emph{ebay.de}. %

\subsection{Continuous Monitoring}
\label{sec:continous_monitoring}
\Gls{ssomon} runs multiple scans in sequence to monitor the \gls{sso} landscape continuously.
It then compares the results and presents their differences.
There are two ways to continuously monitor the \gls{sso} landscape.
Both of them require an initial landscape analysis run, which includes the aggregation of login page candidates.

\paragraph{Punctual Monitoring}
To determine the \gls{sso} landscape at a current timestamp, the analyst creates a continuous monitoring job and selects the first scan as a basis.
The next scan uses the login pages candidates from the first scan.
If the supported \glspl{idp} change, e.g., an \gls{idp} was added or removed, %
the continuous monitoring job will rerun a full detection of login pages and \glspl{idp}.
Such scans can be repeated arbitrarily and represent the \gls{sso} landscape at a specific time.

\paragraph{Constant Monitoring}
To monitor the \gls{sso} landscape constantly, the analyst creates a constant monitoring task.
\gls{ssomon} automatically uses free workers to update the \gls{sso} landscape continuously.
This way, a large corpus of sites can be monitored fully automated over time.
This approach provides an always up-to-date landscape.

\subsection{The Tranco Top 10k \glsentryshort{sso} Landscape}
\label{sec:landscape_results}

\begin{table}[t]
\centering
\resizebox{\columnwidth}{!}{%
\begin{tabular}{c|cc|cc|cccc}
    \toprule
    \multirow{2}{0.5cm}{\textbf{\gls{idp}}} & \multicolumn{1}{c}{\textbf{SSO}} & \multirow{2}{1.0cm}{\textbf{Broken}} & \multicolumn{2}{c|}{\textbf{Protocol}} & \multicolumn{4}{c}{\textbf{Flows}} \\
                & \textbf{Logins} & & \gls{oauth} & OIDC    & Code & Hybrid & Implicit & N/A \\
    \midrule
    \faGoogle   &   1,399 & 98 & 667 & 634 & 946 & 43 & 276 & 36  \\
    \faFacebook &   1,150 & 73 & 1,068 & 9  & 586 &  0   &  314  &  177  \\
    \faApple    &   471   & 38 & 212 & 221 & 236 & 197 &  0    &    0\\
    \midrule
    $\sum$      &   3,020 & 209 & 1,947 & 864 &  1,768   &  240  &  590  & 213  \\
    \bottomrule
\end{tabular}
}
\caption{SSO-Monitor automatically found \num{3,020} SSO logins on the Tranco \num{10k} websites but only \num{2,811} could be further analyzed due to technical constraints.
\Gls{oauth} (\num{69\%}) and the code flow (\num{63\%}) are predominantly used in the wild.}
\label{tab:tranco10k_landscape}
\end{table}

\paragraph{Landscape Overview}
With \gls{ssomon}, we ran the SSO detection process on the Tranco \num{10k} list from \num{July to August 2022}.
We summarize our results in~\autoref{tab:tranco10k_landscape}.
Out of \num{10k} websites, \num{1,389} sites (\num{13,9\%}) were not reachable during our analysis --- most likely due to domain name issues, server errors, or downtime.
We excluded these websites from our analysis, leaving a final dataset containing \num{8,611} sites. 
In total, we found \num{1,632} websites (\num{16,3\%}) offering SSO support with at least one of the three \glspl{idp}. Within this group, the most prevalent \gls{idp} is Google, which is supported on \num{1,399} websites (\num{86\%}), followed by Facebook with \num{1150} sites (\num{70\%}).
Apple started its SSO service in 2019 and is now supported by \num{471} sites (\num{29\%}). %
Interestingly, \num{1,040} out of \num{1,632} sites (\num{64\%}) offer support for multiple \glspl{idp} (\faGoogle+\faFacebook+\faApple: \num{348}, \faGoogle+\faFacebook: \num{576}, \faGoogle+\faApple: \num{86}, \faFacebook+\faApple: \num{30}).
Overall, we identified \num{3,020} SSO logins out of which \num{209} could not be further analyzed. %
Therefore, we excluded them, resulting in a total set of \num{2,811} \gls{sso} logins.
\paragraph{Automated \gls{sso} Discovery}
The SSO discovery reveals that \gls{oauth}, which is designed for authorization, is still the preferred protocol with \num{1,944} \gls{sso} logins (\num{69\%}). In comparison, \gls{oidc}, which allows authentication, is only used in \num{864} \gls{sso} logins (\num{31\%}). In total, \num{590} \gls{sso} logins still use the deprecated implicit flow for authentication. Interestingly, Apple does not use implicit flows. They may not support legacy flows because their SSO service was released quite recently. %

\paragraph{Keyword vs. Image}
Out of the \num{3,020} discovered SSO logins, \num{2,825} were discovered by the keyword-based approach (\autoref{sec:keyword_recognition}) and \num{195} were additionally found by the subsequent image analysis (\autoref{sec:logo_recognition}). Note that the image analysis is only applied if the keyword-based analysis was not successful. %

\paragraph{Identification of Unrelated Logins}
The automated discovery of login pages with search engines produced promising results. 
However, we discovered a low rate of \glspl{fp} in the landscape analysis. %
We used the operator \code{site:domain.com} in the search query to ensure that the discovered login pages are on the same (sub)domain of the Tranco entry.
Nevertheless, pages can redirect to other domains.
For example, the login page \code{photoshop.com/login} redirects to \code{auth.services.adobe.com}.
This redirect can lead to \glspl{fp} if a site redirects to another site that supports \gls{sso}. 
To better understand the problem, we looked into the first \num{500} analysis results.
We identified that \num{3\%} of the sites with \gls{sso} belong to a domain different from its original Tranco entry.
Most of these cases are redirects to Twitter profiles. 
Since Twitter supports \gls{sso}, it is identified as a login page for that domain.

\paragraph{Continues Monitoring}
We compared our latest \gls{sso} landscape results to a different run performed earlier in April 2022. 
As shown in \autoref{tab:diff_cont_mon}, the comparison shows a tremendous variation in \gls{sso} support over four months.

\begin{table}[t]
\centering
\begin{tabular}{r|ccc|c}
\toprule
 \emph{April vs. July}       & \multicolumn{3}{c|}{SSO Logins} & Websites \\
        \emph{2022} & \faGoogle & \faFacebook & \faApple   & %
        \textsupsub{~\faGoogle}{\faFacebook\faApple} \\
\midrule
Added   & \num{152} & \num{144}   & \num{59}   & \num{186} \\
Removed & \num{158} & \num{125}   & \num{57}   & \num{166} \\
\bottomrule
\end{tabular}
\caption{
\gls{sso}-Support Variance in 4 Months.
In total, \num{186} websites added support for at least one provider. %
E.g., \num{152} websites added Google login support, while \num{158} removed it.
}
\label{tab:diff_cont_mon}
\end{table}

%% file: sections/60_security.tex
\section{Automatic Evaluation: \glsentryshort{sso} Security}
\label{sec:security}

\subsection{Methodology}
The \gls{sso} security evaluation consists of three steps, see \autoref{fig:sso_monitor}, and takes the landscape analysis results as input.

\paragraph{Scripted Browser Preparation}
The browser preparation script creates a fresh browser profile with an active \gls{idp}-session and 4 pre-installed browser extensions. We
\begin{inparaenum}
\item use the \emph{i don't care about cookie} extension\footnote{\url{https://www.i-dont-care-about-cookies.eu/}} to automatically remove cookie banners,
\item develop \gls{pm} and \emph{Fragment} extensions to make \glspl{ibc} visible to \gls{ssomon}, and
\item develop \emph{Auto Consent Extension (ACE)} to automatically grant \gls{enduser} consent and automate the \gls{idp} login. %
\end{inparaenum}

\paragraph{SSO Login Execution}
For each \gls{idp}, we visit the login pages in a Selenium-driven Chrome with the appropriate profile.
Next, we use the coordinates from our landscape analysis to click on the \gls{sso} button. %
If clicking the original coordinates does not lead to the expected \gls{idp}, we restart the \gls{sso} detection.
We capture the \gls{http} traffic and \gls{ibc} in HAR files to finally analyze them.
Due to our browser extensions and the authenticated \gls{idp} sessions, this step is fully automated.

\subsection{Security Analysis}

\paragraph{Test Selection}
We selected our security tests according to the following criteria to match our methodology:

\begin{compactenum}
    \item \emph{Passive} tests that do not involve any active parameter manipulation.
    \item Tests detecting faulty behaviour in \emph{\glspl{client}}. We do not investigate \glspl{idp}.
    \item Tests visible in the \emph{\gls{fc}}. Since the \gls{bc} is inaccessible with our approach, we excluded it.
\end{compactenum}

Given the above criteria, we chose the security tests from consolidated security recommendation documents.
\citet{lodderstedtOAuthSecurityBest} from the IETF OAuth working group collaborate with researchers for consolidating current SSO issues and their best practices in a single document.
Since the document is mostly specific for OAuth, we extended our test set with security recommendations section in the \gls{oidc} core specification~\cite{openidConnect}.
We implemented tests that detect the requirements resulting in the following attacks:

\begin{compactitem}
    \item Obsolete Flows, Access Token Disclosure, Implicit Flow Threats~\cite{lodderstedtOAuthSecurityBest,openidConnect}. Test: \gls{access_token} in \gls{fc}.

    \item Open Redirect on the \gls{c}~\cite{lodderstedtOAuthSecurityBest}. Test: Find nested URLs in the \gls{redirect_uri} using \gls{spar}.
    \item CSRF Vulnerability~\cite{lodderstedtOAuthSecurityBest,openidConnect} with state, PKCE, nonce. Test: Missing parameters or insufficient entropy identified with \gls{spar}.

    \item Secret Leakages~\cite{lodderstedtOAuthSecurityBest}. Test: Check HTTP Referer Headers, Tokens in Browser History, or \gls{client_secret} visible in \gls{fc} identified with \gls{spar}.

    \item HTTPS only requests~\cite{lodderstedtOAuthSecurityBest,openidConnect}. Test: Checking all requests for TLS.

    \item Authorization Code Injection, Token Substitution, Access Token Injection~\cite{lodderstedtOAuthSecurityBest,openidConnect}. Test: Missing parameters (PKCE, \gls{nonce}, \gls{at_hash}).

    \item Token Manufacture/Modification~\cite{openidConnect}. Test: \gls{id_token} in \gls{fc} signed with symmetric key.

\end{compactitem}

\noindent We added tests to identify the following protocol violations:
\begin{compactitem}
    \item Protocol Mix-Up: The \gls{c} starts \gls{oauth} but the \gls{idp} switches to \gls{oidc}.
    \item Flow Mix-Up: The \gls{c} started a flow that does not match the \gls{idp}'s returned flow.
\end{compactitem}

Although our test set is distinctive, more tests matching our criteria can simply be added because of our methodology.

\paragraph{Analysis Methodology}
We implemented an \emph{HAR-Analyzer} module to evaluate the recorded HARs.
First, it loads the HAR data into an in-memory graph data model by using a standard HAR parser library\footnote{\url{https://github.com/sdstoehr/har-reader}}.
Next, it extracts the semantic information related to \gls{oauth} and \gls{oidc}, and enriches the model with this information.
For locating the relevant \glspl{lreq} and \glspl{lres}, it scans for data that was sent from a \gls{c} to the \gls{idp} and contains dedicated SSO parameters, for example, \texttt{client\_id} and \texttt{redirect\_uri}.
Then, HAR-Analyzer investigates all further parameters that the \gls{c} website and the \gls{idp} exchange for detecting security issues.
For this purpose, we used our \gls{spar} approach that we describe below.
Finally, HAR-Analyzer produces a report containing the parameters of the SSO messages, the inferred results, and an aggregation of the relevant data.

\begin{figure}
    \includegraphics[width=\linewidth]{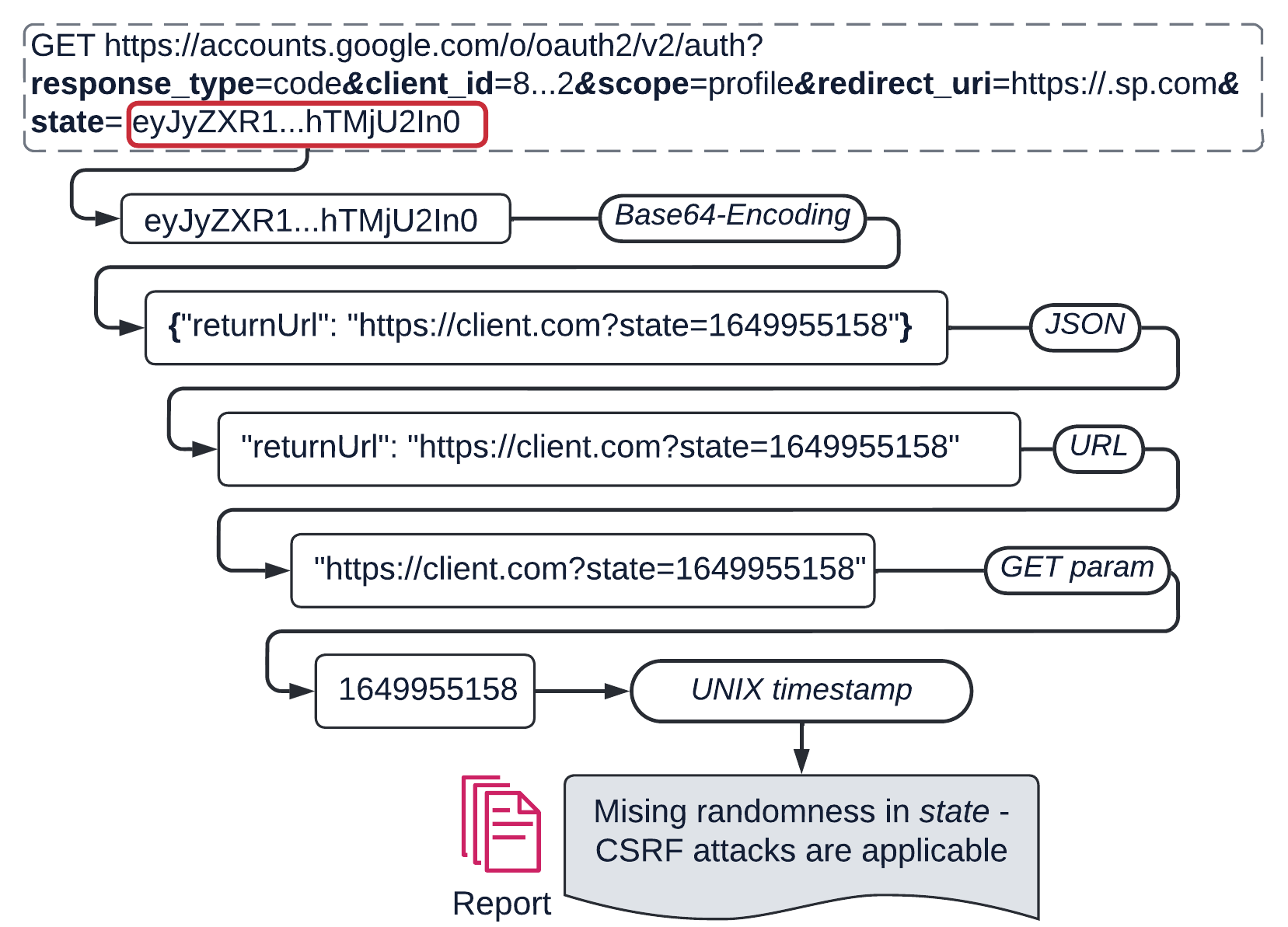}
    \caption{The \gls{spar} approach parses all HTTP parameters until no further data structure can be extracted. Therefore, we can identify deeply nested values in HTTP parameters, so that further security analyses are possible.}
    \label{fig:smartparams}
\end{figure}

\paragraph{\gls{spar}}
\label{sec:smartparams}
We need to inspect structured data for the fully automated security analyses without knowing the exact format.
For example, \gls{oauth} typically uses a random \gls{state} parameter to protect against CSRF attacks.
However, its randomness can be deeply nested in a structured parameter.
The in-depth analysis of this parameter is the core idea of \gls{spar} and depicted in \autoref{fig:smartparams}.
A \glsentryuserii{spar} starts with an initial value, usually a string, and provides a set of decoded or parsed values like URL, WWW Form Encoded, JSON, JWT, and more.
Another \glsentryuserii{spar} object is created for each decoded value, forming a searchable tree.
The \gls{spar} processing step is applied to all gathered requests and responses. Furthermore, the collected \gls{sso} parameters %
are the foundation for further security analyses. %

\subsection{The Tranco Top 10K Security Results}
\label{sec:security_landscape}
In total, we analyzed \num{131} GB of HAR-Files.
It shows that \num{282} (\num{10\%}) of the \gls{sso} executions failed.
We manually investigated these cases and figured out that \num{219} (\num{8\%}) are caused by faulty \gls{client} configurations, e.g., unregistered \gls{redirect_uri}s.
Only \num{63} (\num{2\%}) \glspl{client} executed \gls{sso} successfully.
Due to the long running scanning process, these sites may not be reachable at the time of execution.

We define six security threats systematized in two categories: potential security issues and vulnerabilities.
Potential security issues define four misconfigurations violating either the specification or the security best practices.
These misconfigurations do not necessarily mean that the implementation is vulnerable.
It is a clear indicator that security analysts should provide further investigations.
The second category defines two vulnerabilities, which are considered critical: \gls{csrf} and secret leakage.
Even though the specification and previous researches address \gls{csrf} vulnerabilities clearly, the number of found \gls{csrf} vulnerabilities is surprisingly high.

\newcolumntype{C}[1]{>{\centering\let\newline\\\arraybackslash\hspace{0pt}}m{#1}}
\begin{table}[t]
    \centering
    \resizebox{\columnwidth}{!}{%
    \begin{tabular}{l|cccc|ccc}
        \toprule
                    & \multicolumn{4}{c|}{\textbf{Potential Security Issues}} & \multicolumn{3}{c}{\textbf{Vulnerabilities}} \\
                    & \multirow{2}{1.2cm}{\centering Obsolete Flows} & \multirow{2}{1.14cm}{\centering Protocol Mix-Up}  & \multirow{2}{1.14cm}{\centering Flow Mix-Up} & \multirow{2}{1.1cm}{\centering Open Redirect} & \multicolumn{2}{c}{CSRF} & \multirow{2}{1.1cm}{\centering Secret Leakage} \\
                    &  & &  &  & Weak & Missing  &     \\
        \midrule
        \faGoogle   & 28 & 21 & 39 & 3 &   172 & 162 & 1 \\
        \faFacebook & 314 & 1 & 3 & 5 &   102 & 270 & 1 \\
        \faApple    & 0 & 193 & 4 & 1 &   63 & 15 & 3 \\
        \midrule
        $\sum$      & 342 & 215 & 46 & 9 & 337 & 447 & 5 \\
        \bottomrule
    \end{tabular}
    }
    \caption{We define four potential security threats violating the specifications or the current best practices. We also automatically discover two critical vulnerabilities based on the \gls{spar}-analysis.}
    \label{tab:security_analysis_overview_table}
\end{table}

\paragraph{Obsolete Flows}
Concerning security, we need to highlight the usage of the deprecated implicit flow.
Websites must avoid the transmission of the \gls{access_token} in the \gls{fc} since it increases the risks of its theft~\cite{I-D.ietf-oauth-security-topics}.
As shown in \autoref{tab:security_analysis_overview_table}, \num{342} flows are still implicit or the \gls{c} uses the hybrid flow with the \gls{access_token} in the \gls{fc}.
This makes implicit the second most used flow after the code flow.

\paragraph{Protocol Mix-Up}
In \autoref{tab:security_analysis_overview_table}, we analyzed the violations concerning the started \gls{sso} protocol -- \gls{oauth} vs. \gls{oidc}.
The \gls{oidc} conformance is checked for all recordings which contain indicators that the \gls{c} uses \gls{oidc} rather than plain \gls{oauth}.
Deviations from the specification are noticed whenever the required \texttt{openid} scope value is not present but \gls{id_token} is returned.
This behavior can be primarily attributed to Apple (\num{193}), as the websites don't initialize the scope parameter correctly,
Even though this is a clear violation of the specification, the security implications are unclear and depend on the corresponding implementation.
The risk occurs if unsolicited tokens are processed, but the corresponding verification is skipped.

\paragraph{Flow Mix-Up}
Additionally, we identified threats in which a \gls{c} initiates the code flow but the \gls{idp} proceeds with the implicit flow, that is, it returns an \gls{id_token} or \gls{access_token}.
The same check is performed when a token is requested and an \gls{id_token} or \num{\gls{code}} is additionally returned.
\autoref{tab:security_analysis_overview_table} shows, that \num{46} flow mix-up occurred -- mainly on Google.
We assume that this misconfiguration occurs when the \gls{idp} is configured to support a specific flow per Client while ignoring the \gls{response_type} in the \gls{lreq}.
Such behavior facilitates cut-and-paste attacks and makes the redemption of stolen tokens easier~\cite[Section 4.6]{I-D.ietf-oauth-security-topics}.

\paragraph{Open Redirect}
Open redirect attacks provide a means to redirect the victim to an attacker-controlled location. %
In \gls{oauth}, the \texttt{redirect\_uri} parameter is allowed to contain further query parameters such as URLs.
If the URL contained in the \texttt{redirect\_uri} is followed without any verification on the \gls{c}, then the \gls{c} is susceptible to a open redirect flaw.
In our analysis, we used \gls{spar} for deep inspection of the URLs encoded in query parameters, for instance, in \texttt{redirect\_uri}.
Our findings are shown in \autoref{tab:security_analysis_overview_table} and confirm that the security best practices are well applied.
In \num{9} login \gls{sso} flows, we identified nested URL parameters.
This occurrence does not necessarily indicate a vulnerability as (1) the \gls{c} might have further restrictions implemented and (2) the \gls{idp} can strictly validate the \texttt{redirect\_uri} against an allowlist.
Manual verification revealed, that one \gls{client} redirected to a manipulated open redirect URL.
Other \glspl{client} aborted the \gls{sso} process when they received the \gls{authresp} or, in fact, the \gls{idp} recognized the manipulated \gls{redirect_uri} and denied the login.

\paragraph{\glsentrylong{csrf}}
\Gls{csrf} is a common web attack that enables an attacker to execute sensitive operations in the context of a victim.
The only requirement for a \gls{csrf} attack is that the victim visits a link provided by the attacker.
In the context of \gls{sso}, two different \gls{csrf} attack variants are possible:
\begin{inparaenum}
    \item \emph{Session Swapping} aims to log the victim into an account belonging to the attacker, and
    \item \emph{Force Login} is an attack to log the victim into his own account.
\end{inparaenum}
For mitigating the attack, \gls{oauth} and \gls{oidc} rely on two parameters -- \code|state| and \code|nonce| which contains an unpredictable random string.
If both parameters are missing or the values are predictable, then \gls{csrf} attacks can be executed.\footnote{\gls{pkce} can be used instead of \code|state| and \code|nonce|. In this case, we do not classify the website as vulnerable.}
The number of vulnerable logins is surprisingly high -- \num{447} logins do not use any of these parameters.
In addition, we could find \num{337} websites implementing an insufficient \gls{csrf} protection with low entropy ($\leq 96$ bit), based on our \gls{spar} analysis.

\paragraph{Secret Leakage}
In \gls{sso}, the \gls{c} authenticates on the \gls{idp} via the \gls{client_secret}.
If attackers obtain the \gls{client_secret}, \emph{Client Impersonation} attacks can be executed~\cite[Section 5.2.3]{RFC6819},\cite{ssoscan}.
In addition, the \gls{id_token} should not be signed with symmetric cryptography when the implicit flow is used~\cite[Section 10.1]{openidConnect}, because the \gls{c} cannot verify the token without leaking the key.
In \autoref{tab:security_analysis_overview_table}, one can see that only \num{5} \glspl{client} leak secrets.
We did not discovered any \glspl{id_token} using symmetric cryptography.
This is motivated by the fact that the \glspl{idp} always sign the \gls{id_token} with RSA or elliptic curves.

\paragraph{Advanced Security Countermeasures}
The IETF aims to reduce the risks against attacks to a minimum.
In addition to the standardization of security consideration~\cite{RFC6819} and security best practices~\cite{I-D.ietf-oauth-security-topics}, the IETF developed protection mechanisms applied as extensions on the top of \gls{oauth} and \gls{oidc}.
Such protection mechanisms are \gls{pkce}~\cite{RFC7636}, \gls{dpop}~\cite{ietf-oauth-dpop-07}, and mTLS~\cite{RFC8705} .
While \gls{pkce} can be detected by analyzing the network traffic, \gls{dpop} and mTLS are executed only in the back-channel communication between the \gls{c} and the \gls{idp}. Although, it was originally designed for native applications, \gls{pkce} can be used to prevent authorization code injection attacks \cite{lodderstedtOAuthSecurityBest}. Our analysis discovered \num{23} \glspl{client} implementing \gls{pkce} -- \num{14} (\faGoogle), \num{3} (\faFacebook), and \num{6} (\faApple).
Additionally, \gls{oidc} defines the \gls{nonce} parameter and therefore provides another way to prevent authorization code injection attacks. However, the usage of the \gls{nonce} parameter in general is quite low. In total, \num{493} \gls{client}s requested an \gls{id_token} explicitly in the \gls{response_type} but only \num{42} of them created a related \gls{nonce} parameter. Also, none of the \gls{client}s requesting an \gls{id_token} use \gls{pkce}.

Similar to the \gls{code}, the \gls{at} can also be a target of an injection attack. Unfortunately, there are no ways to detect such an attack on the \gls{oauth} protocol level. However, \gls{oidc} benefits from the \gls{id_token} which contains the \gls{at_hash}. Hence, the \gls{at} can be validated on usage time. From \num{39} \gls{client}s which requested an \gls{id_token} and \gls{at} all \gls{id_token} included an \gls{at_hash}.

\paragraph{HTTPS Only Request}
As stated in the \gls{oauth} security best practices \cite{I-D.ietf-oauth-security-topics}, each authorization responses must not be transmitted over unencrypted network connections. However, we found \num{14} \gls{sp}s which try to set an unencrypted plain http \gls{redirect_uri} -- 5 \faGoogle, 9 \faFacebook). Additionally, \num{4} of them received a valid authorization response. Interestingly, Facebook blocks plain http \gls{redirect_uris} while all authorization responses were send by Google. The only exception for using unencrypted conections are native clients that use loopback interface redirection. None of the aforementioned cases apply to this exception. When looking at the whole http traffic, \num{136} \gls{sp}s try to send unencrypted http requests from which \num{32} also get a valid response.

%% file: sections/70_privacy.tex
\section{Automatic Evaluation: \glsentryshort{sso} Privacy}
\label{sec:privacy}

\subsection{Methodology}
The SSO privacy evaluation consists of three steps as depicted in ~\autoref{fig:sso_monitor}:
\begin{inparaenum}
\item For each automated privacy analysis conducted by SSO-Monitor, we use a script to create a fresh browser profile with an active IdP session. We created profiles for Google, Facebook, and Apple.
\item We use the browser profile from step (1) to load each website from the Tranco 10k list supporting the top three IdPs in a Chrome browser. Again, we use Selenium for automation and to capture traffic in HAR files.
\item Our HAR Analyzer module automatically analyzed the HAR files created in step (2) for possible privacy leaks.
\end{inparaenum}

\paragraph{Scripted Browser Preparation}
The scripted browser preparation consists of two parts.
First, we reused the three profiles we used during the security analyses.
This reusing is necessary since we can be confident that each \gls{idp}-account was used previously to sign in to a certain \gls{client}.
Without this reusing, we would not know whether the account provided consent for this particular \gls{client} to the \gls{idp}.
We call them the \emph{consent-given profiles}.
Second, we create one new browser profile per \gls{idp} using a fresh \gls{idp}-account. 
With this, we can be confident that these accounts have \emph{never} logged into the \gls{client} before.
Thereby, no consent is given and 
we call them \emph{no-consent profiles}.
Note that the profiles only contain \gls{idp}-related cookies. Therefore, we ensure that a user is logged out on all \glspl{client}.
In summary, our privacy analyses use, thereby, six different browser profiles.

\paragraph{Client Website Visit}
For each website that our landscape analyses provides, we start visiting them.
For each supported \gls{idp} on the website, we open the start page\footnote{In contrast to the security analyses, during which we visited the login page.} twice with the corresponding browser profiles.
The first time with the \emph{consent-given profile}, the second time with the \emph{no-consent profile}. 
In both cases, we interact with the website by clicking on random links which do not claim to require or start any authentication and pressing some keys (e.g., PageUp).
In summary, if a website supports all three \glspl{idp} (Apple, Google, Facebook), we visit the start page six times, once with each browser profile, and interact with that page.
\gls{ssomon} records the traffic of each visit.

\subsection{Privacy Analysis}

\paragraph{Test Selection}
For the selection of privacy tests in \gls{sso}, there exists no document summarizing such issues in contrast to security tests.
Thereby, our privacy analyses detect whether a website visitor is authenticated in the background, for instance, without explicitly clicking a sign-in button.
We additionally search for any identity-related information leaks to the \gls{client}.
With our \emph{consent-given profile}, the \gls{client} can -- in theory -- log in the user in the background.
Our analyses reveal that this is abused in \num{199} cases.
In contrast, the \emph{no-consent profile} should protect users from this behavior since they never agreed to share their identity with the \gls{client}.
In both cases, an automatically created login attempt, created by the \gls{client}, may allow the IdP to track users secretly.
Considering users' privacy, we raise the following research questions:
\begin{inparaenum}[(1)]
    \item Do websites exchange \gls{sso} messages revealing privacy information without users' consent?
    \item Do the \glspl{idp} provide a sufficient level of protection against \emph{honest but curious \glspl{client}}? In our threat model, an honest but curious \gls{client} acts according to the protocol and establishes a trust relationship with the \gls{idp}. 
    \Glspl{client} can easily gain this relationship because \glspl{idp} support the registration of arbitrary \glspl{client}.
\end{inparaenum}

\paragraph{Leakage Channels}
A login attempt signals the beginning of the \gls{sso} authentication.
The \gls{client} initiates the protocol and sends a \gls{lreq} to the \gls{idp} along with the \gls{idp}'s session cookie.
Each message can disclose different private information.
Our evaluation considers only messages as a leak if the user has not explicitly started any login.
We classify \emph{leakage channels} in two different categories: login attempt and token exchange.

\paragraph{Login Attempt Leak (LAL): Privacy Leak to \glspl{idp}}
The first leakage occurs if the \gls{lreq} is sent without the user actively navigating to the login page at the \gls{client}.
If the user is authenticated to the \gls{idp}, the \gls{idp} learns which website the user is currently navigating.

\paragraph{Token Exchange Leak (TEL): Privacy Leakage to \glspl{client}}
The second leakage targets the \gls{lres}.
It contains the user's identity, for example, the email.
If the \gls{idp} issues the \gls{lres} without any consent, the \gls{client} learns the user's identity.
Therefore, the user's identity is entirely revealed to both \gls{sso} parties.

\paragraph{Cookie-based vs. SSO-based Privacy Leaks}
In contrast to cookie-based privacy leaks, our findings are more invasive.
Usually, the user visits a \gls{client} and actively decides to click on the sign-in button. 
After successful authentication, the browser stores the session cookies. If the user does not clear the cookie store, the website can re-identify the user based on the cookies.
In \gls{sso} privacy evaluation, the user visits the \gls{client} and the \gls{idp} automatically returns user-identifiable tokens. 
Even if no cookies for the \gls{client} are stored,
the website can still identify the user at any time.
Consider a user starting the browser for the first time. 
The \gls{enduser} authenticates to an \gls{idp}, for example, to synchronize the browser settings.
If the user afterward visits the \gls{client}, the identity automatically leaks to the \gls{client}.
We claim our privacy leaks as novel.
To our best knowledge, no previous work has investigated such token leaks in an automated manner.

\subsection{The Tranco Top 10K Privacy Results}

We evaluated all websites of the Tranco top 10k list supporting logins with the top three \glspl{idp} (Facebook, Google, and Apple) and discovered multiple privacy leakages, which we discuss in this section. In total, we analyzed \num{135 GB} of HAR-Files. 
We summarize the results in \autoref{tab:privacy}.
Note that among \num{3,020} detected SSO logins, \num{24} privacy analyses are missing as the related websites failed to load during the run.
We determined two categories with our pre-generated profiles: no consent given and consent given.
\begin{table}[t]
\centering
\begin{tabularx}{\linewidth}{XcYYYY}
\toprule
 & & \multicolumn{2}{>{\centering\hsize=2\hsize\linewidth=\hsize}X}{\textbf{No-Consent Profiles}} & \multicolumn{2}{>{\centering\hsize=2\hsize\linewidth=\hsize}X}{\textbf{Consent-Given Profiles}} \\
IdP          & SSO Logins & LAL & TEL & LAL & TEL  \\ 
\midrule
\faGoogle    & 1,387   & 161  & 0   & 160  & 22  \\
\faFacebook  & 1,143   & 39  & 0   & 39  & 20  \\
\faApple     & 466 & 0   & 0   &  0  & 0  \\ 
\midrule
$\sum$       & 2,996   & 200  & 0   & 199  & 42 \\
\bottomrule
\end{tabularx}
\caption{We discovered that in \num{200} of \num{2,996} (\num{7\%}) \gls{sso} logins, the login is initiated automatically by visiting the \glspl{client}' starting page. On \num{42} (\num{1\%}) of them, the \gls{idp} automatically sends authentication tokens and reveals the user's identity.
Our leakage findings must be seen as lower boundaries.} %
\label{tab:privacy}
\end{table}

\paragraph{Case 1) No-Consent Profiles}
We identified \num{200} \glspl{lreq} being sent transparently to the Google and Facebook \glspl{idp}.
Interestingly, we did not observe automatic login attempts to Apple, possibly because Apple enforces a user to consent on every login attempt.
Also, a positive result is that none of the \glspl{idp} generate authentication tokens automatically in this category.
This is the correct and expected behavior, since the user should consent at least the first time when an authorization by the \gls{client} takes place.

\paragraph{Case 2) Consent-Given Profiles}

We observed almost the same amount of login attempts for Google as in the previous category, which is not surprising since the \gls{client} does not know a-priory whether a user is authenticated to any \gls{idp}. The one missing LAL can be attributed to the client not being available at the time of the scan. %
Overall, we discovered that in \num{42} cases, the \glspl{idp} automatically generates authentication tokens and send them to the \gls{client} in the \gls{lres}.
Thus, the \gls{client} observes the user's identity even if the user never consciously started any authentication on the \gls{client}. Regarding Google, all \num{22} found TELs are due to Google's \emph{autologin} feature, where the user automatically gets logged in to the \gls{client}. Although this behavior is not desirable for the users' privacy, these cases may be visible to the user. In contrast, Facebook does not transparently show the login process. Our analysis reveals that \num{20} websites stealthily learn the user's identity by TELs. Hence, it is up to the \gls{client} if they reveal this to the user or not (i.e. via UI).

%% file: sections/80_related_work.tex
\section{Related Work}
\label{sec:rw}

Apart from the systematization of known SSO tools in \autoref{sec:sso_tools}, there is more related work on \gls{sso}.
We divided prior work into three categories to match \gls{ssomon}'s architecture.

\paragraph{\glsentrylong{sso} Landscape}

In \citeyear{alaca2020ComparativeAnalysisFramework}, \citet{alaca2020ComparativeAnalysisFramework} developed a framework to compare protocol designs and evaluate 14 different web \gls{sso} systems, but they did not compare implementations. %
In \citeyear{morkondaExploringPrivacyImplications2021}, \citeauthor{morkondaExploringPrivacyImplications2021} analyzed the Alexa top 500 per country for five countries~\cite{morkondaExploringPrivacyImplications2021}. They categorized user data provided by Google, Facebook, Apple, and LinkedIn, and identified that \glspl{client} request different data for different \glspl{idp}. They analyzed the \gls{lreq}, while we investigated the whole login flow. %

\paragraph{Practical Security Evaluations}

A considerable amount of literature has been published on the security of \gls{oauth} and \gls{oidc} web implementations in the wild.
A significant amount of researcher concentrate on classical web vulnerabilities in the \gls{sso} context, such as \gls{csrf} and \gls{xss}~\cite{Sun2012,li2014security,Shernan2015,li2018mitigating}.
\citet{li2016analysing} investigated specific issues in Google's \gls{oidc} implementation.
\citet{wang2016AchillesHeelOAuth} analyzed \gls{oauth} protocol implementation on various platforms~\cite{wang2016AchillesHeelOAuth}.
\citet{sok_oidc} identified issues in offical \gls{oidc} libraries.
\citet{Sadqi2020} provided in \citeyear{Sadqi2020} a survey of OAuth relevant threats for web clients.
Recently, \citet{saito2021ComparisonOAuthOpenID} assessed the implementation of social logins on 500 American websites and compared their results to Japanese websites (\citeyear{saito2021ComparisonOAuthOpenID}). They found that 76 websites are susceptible to attacks, mainly caused by faulty implementations or insecure design decisions. %
In \citeyear{Liu2021}, \citeauthor{Liu2021} introduced a new threat for \gls{sso} authentication~\cite{Liu2021}. 
In comparison to our work, the authors did not concentrate on protocol flaws and implementation mistakes, but on a design issue of the \gls{sso} ecosystem. They reused abandoned email addresses.

\paragraph{Privacy}

Besides a large corpus of literature concerning \gls{sso} security, a considerable amount of research has focused on privacy issues in \gls{sso}.
Various research groups build new or extend existing \gls{sso} schemes to tackle their privacy concerns~\cite{fett2015spresso,isaakidis2016unlimitid,hammann2020PrivacyPreservingOpenIDConnect,zhang2021passo}.
These issues range from leaking user-specific parameters up to complete identity revelation.
\citet{farooqi2017MeasuringMitigatingOauth} investigated how leaked Facebook tokens are abused. %
\citet{li2020UserAccessPrivacy} systematically analyzed how \glspl{idp} can track \glspl{enduser}' interactions with \glspl{client}. %
Despite previous efforts in improving the privacy of \gls{oauth} and \gls{oidc}, we still do not see these enhancements being implemented. %

%% file: sections/90_futurework.tex
\section{Conclusion \& Discussion}
\label{sec:futurework}

We conclude with lessons learned and present new directions for future \gls{sso} research.

\paragraph{Further IdPs}
Parts of \gls{ssomon}, for instance, the landscape analyses, already support LinkedIn, Microsoft, Twitter, and Baidu.
However, the integration for security tests requires more complex adaptions.
\gls{ssomon} needs support for the \gls{idp} specific consent pages and the automatic login. Also, the browser profile generation is challenging. 
Additionally, when running the analyses, each analyzed \gls{idp} extends the processing time. Therefore, we decided first to analyze the top three \glspl{idp} and provide landscape, security, and privacy insights. Data for more \gls{idp} will be provided in the future.

\paragraph{Developers vs. Specifications}
Any deviation from the specification makes an automated analysis searching for \gls{sso} patterns hard.
As a result, we developed multiple strategies to recognize \gls{sso} flows even if they are not compliant to the specification.
We also observed a disregard for the security best practices.
Security problems that are well-studied and documented still exist.

\paragraph{Real-World Security Best Practice Compliance}
Our research reveals for one more time that existing security best practices are often ignored and not implemented in the Tranco \num{10k}.
The question arises of how this situation could be improved.
For example, the deployment of TLS on websites has tremendously improved since browsers penalize websites not supporting TLS.
Similarly, this would be possible for \glspl{idp}. For example, they could drop \glspl{lreq} not following best practices.
Future research should concentrate on how secure by default configurations can be deployed more efficiently.

%% file: sections/100_flowchart.tex
\section{\glssc{ssomon}: Flowchart}
\label{sec:ssomon_flowchart}

\begin{figure}[t]
    \centering
    \includegraphics[width=\linewidth]{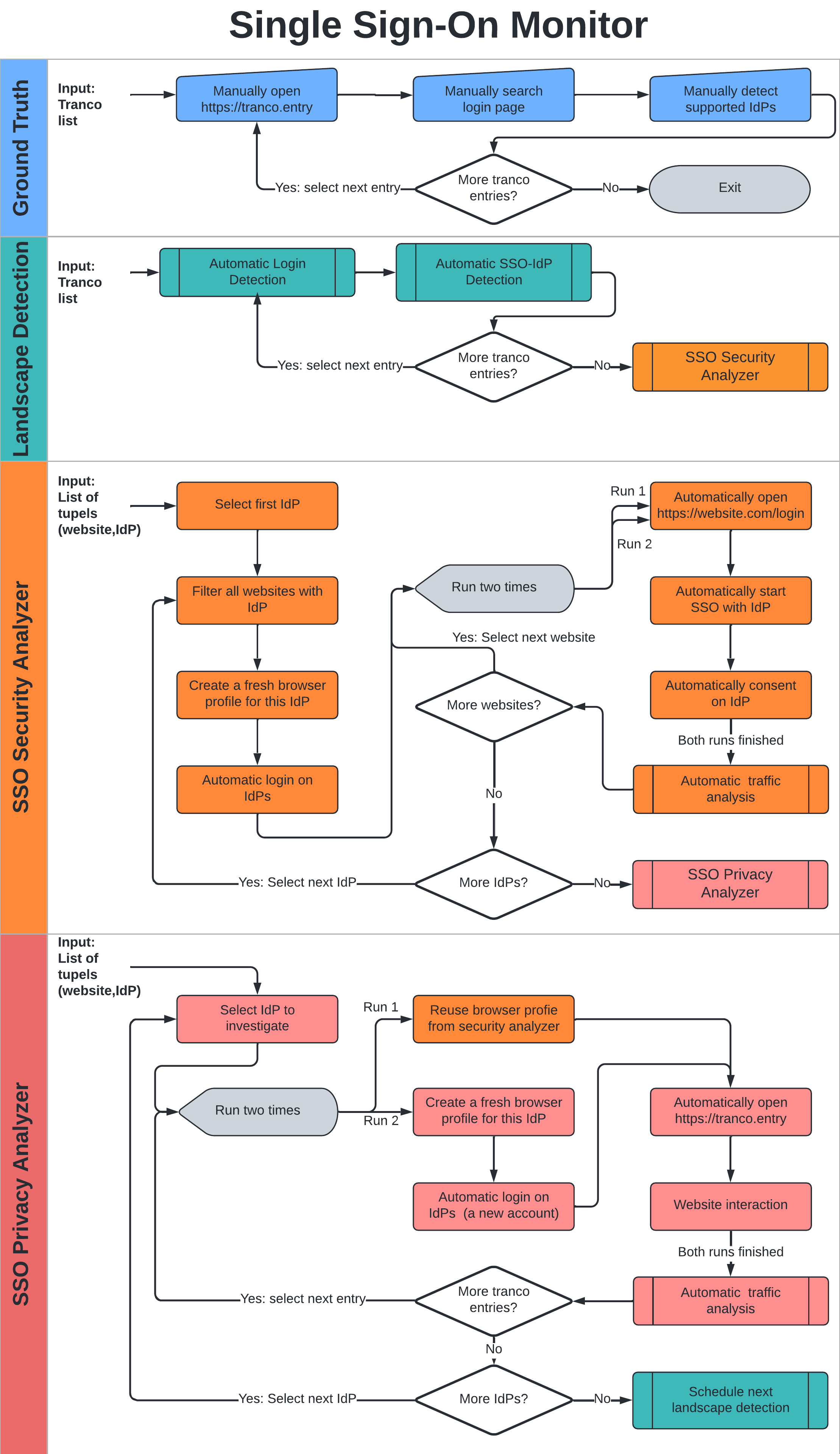}
    \caption{\gls{ssomon}'s workflow}
    \label{fig:sso_monitor_detailed_flowchart}
\end{figure}

\autoref{fig:sso_monitor_detailed_flowchart} enriches \gls{ssomon} as depicted in \autoref{fig:sso_monitor}) with more details.

%% file: sections/120_attack_list.tex
\section{Complete Attack List}

The following lists shows all attacks described in OAuth Security Topics from July 2022 (Draft Version 20) and the OIDC core spec 1.0.

\begin{enumerate}
    \item \textbf{Obsolete Flows, Access Token Disclosure, Implicit Flow Threats}
    \begin{description}
        \item [Definition] Security Topics, Sec. 2.1.2; OIDC Core, Sec. 16.5 and 16.16
        \item [Implementation note] Check for implicit flow.
    \end{description}
    
    \item \textbf{Open Redirect}
    \begin{description}
        \item [Definition] Security Topics, Sec. 4.10
        \item [Implementation note] Find nested URLs in the \gls{redirect_uri} using \gls{spar}.
    \end{description}
    
    \item \textbf{CSRF}
    \begin{description}
        \item [Definition] Security Topics, Sec. 4.7
        \item [Implementation note] Check if PKCE and/or state, nonce is used.
    \end{description}
    
    \item \textbf{Secret Leakage}
    \begin{description}
        \item [Definition] Security Topics, Sec. 4.2 and 4.3
        \item [Implementation note] Check if secrets are present in any request or referrer header and check if a referrer policy is set or PKCE is used.
    \end{description}
    
    \item \textbf{TLS only}
    \begin{description}
        \item [Definition] Security Topics, Sec. 2.6; OIDC Core, Sec. 16.1 and 16.17
        \item [Implementation note] Check that all requests use TLS.
    \end{description}
    
    \item \textbf{Authorization Code Injection, Token Substitution}
    \begin{description}
        \item [Definition] Security Topics, Sec. 4.5; OIDC Core, Sec. 16.11
        \item [Implementation note] Check if PKCE or nonce is used.
    \end{description}
    
    \item \textbf{Access Token Injection, Token Substitution}
    \begin{description}
        \item [Definition] Security Topics, Sec. 4.6; OIDC Core, Sec. 16.11
        \item [Implementation note] No check for OAuth, in OIDC check at\_hash in id\_token. Only applies to implicit.
    \end{description}

    \item \textbf{307 Redirect}
    \begin{description}
        \item [Definition] Security Topics, Sec. 4.11
        \item [Implementation note] Check if 307 Redirect is used for Authorization Responses.
    \end{description}
    
    \item \textbf{Token Manufacture/Modification}
    \begin{description}
        \item [Definition] OIDC Core, Sec. 16.3
        \item [Implementation note] Check that id\_token is signed with asymmetric key.
    \end{description}
    
    \item \textbf{Idp Mixup}
    \begin{description}
        \item [Definition] Security Topics, Sec. 4.4; OIDC Core, Sec. 16.15
        \item [Implementation note] Not implemented as this requires an active attack.
    \end{description}
    
    \item \textbf{Need for Signed/ Encrypted Requests}
    \begin{description}
        \item [Definition] OIDC Core, Sec. 16.20 and 16.21
        \item [Implementation note] Not implemented as requirements of the clients are not known.
    \end{description}
    
    \item \textbf{Insufficient Redirect URI Validation}
    \begin{description}
        \item [Definition] Security Topics, Sec. 4.1
        \item [Implementation note] Not implemented as patterns can only be found by actively trying different redirect\_urls.
    \end{description}

    \item \textbf{PKCE Downgrade}
    \begin{description}
        \item [Definition] Security Topics, Sec. 4.8
        \item [Implementation note] Not implemented as an active attack is necessary to detect whether countermeasures are in place.
    \end{description}

    \item \textbf{Access Token Phishing by Counterfeit Resource Server}
    \begin{description}
        \item [Definition] Security Topics, Sec. 4.9.1
        \item [Implementation note] Not implemented as attack assumes control some control over the client over secondary configuration.
    \end{description}

    \item \textbf{TLS Terminating Reverse Proxies}
    \begin{description}
        \item [Definition] Security Topics, Sec. 4.12
        \item [Implementation note] Not implemented as it is not detectable without trying certain headers in an active attack.
    \end{description}

    \item \textbf{Refresh Token Protection, Lifetimes of Access Tokens and Refresh Tokens}
    \begin{description}
        \item [Definition] Security Topics, Sec. 4.13; OIDC Core, Sec. 16.18
        \item [Implementation note] Not implemented as it can only be checked when auditing client and AS implementation or configuration parameters.
    \end{description}

    \item \textbf{Client Impersonating Resource Owner}
    \begin{description}
        \item [Definition] Security Topics, Sec. 4.14
        \item [Implementation note] Not implemented as resource server access is needed.
    \end{description}

    \item \textbf{Clickjacking}
    \begin{description}
        \item [Definition] Security Topics, Sec. 4.15
        \item [Implementation note] Not implemented as this only affects the AS, not clients.
    \end{description}

    \item \textbf{Server Masquerading}
    \begin{description}
        \item [Definition] OIDC Core, Sec. 16.2
        \item [Implementation note] Not implemented as this requires access to server to server communication.
    \end{description}
    
    \item \textbf{Access Token Disclosure}
    \begin{description}
        \item [Definition] OIDC Core, Sec. 16.4
        \item [Implementation note] Check that no implicit flow is used.
    \end{description}
    
    \item \textbf{Server Response Disclosure}
    \begin{description}
        \item [Definition] OIDC Core, Sec. 16.5
        \item [Implementation note] Check that no implicit or hybrid flow is used.
    \end{description}
        
    \item \textbf{Access Token Redirect}
    \begin{description}
        \item [Definition] OIDC Core, Sec. 16.8
        \item [Implementation note] Not implemented as mitigation can only be checked server side.
    \end{description}

    \item \textbf{Token Reuse}
    \begin{description}
        \item [Definition] OIDC Core, Sec. 16.9
        \item [Implementation note] Not implemented as this requires an active attack.
    \end{description}

    \item \textbf{Eavesdropping or Leaking Authorization Codes (Secondary Authenticator Capture)}
    \begin{description}
        \item [Definition] OIDC Core, Sec. 16.10
        \item [Implementation note] Not implemented as the attack vector is compromised User Agent.
    \end{description}

    \item \textbf{Timing Attack}
    \begin{description}
        \item [Definition] OIDC Core, Sec. 16.12
        \item [Implementation note] Not implemented as this requires a timing analysis of valid and erroneous responses.
    \end{description}

    \item \textbf{Other Crypto Related Attacks}
    \begin{description}
        \item [Definition] OIDC Core, Sec. 16.13
        \item [Implementation note] Not implemented as this requires an active attack.
    \end{description}

    \item \textbf{Signing and Encryption Order}
    \begin{description}
        \item [Definition] OIDC Core, Sec. 16.14
        \item [Implementation note] Not implemented as this is a legal issue which is not security relevant.
    \end{description}
    
    \item \textbf{Issuer Identifier}
    \begin{description}
        \item [Definition] OIDC Core, Sec. 16.15
        \item [Implementation note] Not implemented as this requires a check against the OIDC discovery document.
    \end{description}
    
    \item \textbf{Symmetric Key Entropy}
    \begin{description}
        \item [Definition] OIDC Core, Sec. 16.19
        \item [Implementation note] Not implemented as this requires access to the client's secrets.
    \end{description}
            
\end{enumerate}